\documentclass[%
 aip,
 jmp,%
 onecolumn, %RC
 amsmath,amssymb,
 showkeys,
 preprint,tightenlines,%
]{revtex4-1}

\usepackage{graphicx}% Include figure files
\usepackage{dcolumn}% Align table columns on decimal point
\usepackage{bm}% bold math
\def\today{Submitted July 13, 2017, accepted October 12, 2017}
\def \ccomma{\raise 2pt\hbox{,}} % Le petit livre de TeX page 234
\def \D {\hbox{d}}
\def\LMP{Lett.~Math.~Phys.~}
\def\CRAS{C.~R.~Acad.~Sc.~Paris}
\def\AnnENS{Ann.~\'Ec.~Norm.~}

\def \Rcubec {\mathbb{R}^3(c)}

\def \Pn     {{\rm P_{\rm n}}}
\def \PVI    {{\rm P_{\rm VI}}}
\def \PV     {{\rm P_{\rm V}}}
\def \PIV    {{\rm P_{\rm IV}}}
\def \PIII   {{\rm P_{\rm III}}}
\def \PII    {{\rm P_{\rm II}}}
\def \PI     {{\rm P_{\rm I}}}
\def \qPVI   {{\rm q\hbox{-}P_{\rm VI}}}

\def \diag {\mathop{\rm diag}\nolimits}

\def \mod#1{\vert #1 \vert}
\def \modQ {\mod{Q}}
\def \barQ {\overline{Q}}
\def \fnewq{q}
\def \bfF {{\bf F}}
\def \bfN {{\bf N}}
\def \bfsigma {{\bf \sigma}}
\def \barz {{\bar z}}
\def \barZ {{\bar Z}}

\def \cz {c_{\rm z}}

\def \Dlogtau{\frac{\D}{\D x}\log \tau} % Logarithmic derivative of a tau-function. 
\def \Hqp{\textit{\sl H}} % Notation for H(q,p,x)
\def \HVI {\Dlogtau_{\rm VI}} % ? NOT GOOD. Confusion.

\def \sqrtYone{\sqrt{X(X-1) Y'}} 

\def \fys {Y_{\rm s}}
\def \aT    {a_{\rm T}} % Hamiltonian of Tsegel'nik. Change to H %RC
\def \aX    {a_{\rm X}} % ratio $X/\Omega$
\def \auu{a}
\def \A     {e} % X(X-1)V'+\Theta_X V(V-1) 
\def \Mzero {e_0}
\def \Mone  {e_1}

\def \metric{\upsilon}
\def \matU {\mathbb{U}}
\def \matV {\mathbb{V}}
\def \vectorpsi{{\bf\psi}} % Editor, please make it different from $\psi$

\def \abrfive {r_5}
\def \abrfour {r_4}
\def \abrthree{r_3}
\def \abrtwo  {r_2}

\begin{document}

\preprint{CMLA-2017-07}

\title[Bonnet and $\PVI$]
{
   Generalized Bonnet surfaces and Lax pairs of $\PVI$
}% Force line breaks with \\
%\thanks{Footnote to title of article.}

% AUTHOR HERE

\author{Robert Conte}
 %\altaffiliation[Also at ]{Department.}%Lines break automatically or can be forced with \\
 %\email{Second.Author@institution.edu.}

\affiliation{ 
  \noindent 1.
    Centre de math\'ematiques et de leurs applications,  
\\ \'Ecole normale sup\'erieure de Cachan, CNRS, Universit\'e Paris-Saclay,
\\ 61, avenue du Pr\'esident Wilson, F--94235 Cachan Cedex, France.
\smallskip
\\ \noindent 2.
Department of mathematics,
The University of Hong Kong,
\\ Pokfulam road, Hong Kong.
\smallskip
\\ \noindent E-mail: Robert.Conte@cea.fr
}%

% \homepage{http://www.Second.institution.edu/~Charlie.Author.}
%\affiliation{%
%Second institution and/or address%\\This line break forced% with \\
%}%

\date{\today}% It is always \today, today,

% =======================================================================
% Only the last abstract is kept.

\begin{abstract}
We build analytic surfaces in $\Rcubec$ represented by the most general
sixth Painlev\'e equation $\PVI$ in two steps.
Firstly, the moving frame of the surfaces built by Bonnet in 1867 
is extrapolated 
to a new, second order, isomonodromic matrix Lax pair 
of $\PVI$,
whose elements depend rationally on the dependent variable
and quadratically 
on the monodromy exponents $\theta_j$.
Secondly, by converting back this Lax pair to a moving frame,
we obtain 
an extrapolation of Bonnet surfaces to surfaces with two 
more degrees of freedom.
Finally, we give a rigorous derivation of the quantum correspondence for $\PVI$.
\end{abstract}

% PACS and KEYW HERE

\pacs{
%https://www.aip.org/publishing/pacs/pacs-2010-regular-edition
                        %https://www.aip.org/publishing/pacs/pacs-alphabetical-index		
%02.20.Sv Lie algebras of Lie groups	
\\ 02.30.Hq Ordinary differential equations
\\ 02.30.Ik Integrable systems 
\\ 02.30.Jr Partial differential equations
% (cosmology, general relativity, nonlinear ODEs, quantum gravity)
\\ 02.30.-f Function theory, analysis						
\\ 02.40.Dr Euclidean and projective geometries
\\ 02.40.Hw Classical differential geometry 
}% PACS, the Physics and Astronomy Classification Scheme.

\keywords{\\ Gauss-Codazzi equations;
Bonnet surfaces;
sixth Painlev\'e equation;
Lax pair;
generalized heat equation;
quantum correspondence.}%Use showkeys class option if keyword
															
\maketitle

% =======================================================================
%\vfill\eject
\tableofcontents

\vfill\eject
% ============================================================================
\section{Introduction}
% ============================================================================

{}From the very beginning \cite{Picard1889}, 
two representations have coexisted for the $\PVI$ equation.
The first one, 
\begin{eqnarray}
& & {\hskip -20.0 truemm}
\frac{\D^2 u}{\D x^2}=
 \frac{1}{2} \left[\frac{1}{u} + \frac{1}{u-1} + \frac{1}{u-x} \right] \left(\frac{\D u}{\D x}\right)^2
- \left[\frac{1}{x} + \frac{1}{x-1} + \frac{1}{u-x} \right] \frac{\D u}{\D x}
\nonumber \\ & & {\hskip -20.0 truemm}
+ \frac{u (u-1) (u-x)}{2 x^2 (x-1)^2}
  \left[\theta_\infty^2 - \theta_0^2 \frac{x}{u^2} + \theta_1^2 \frac{x-1}{(u-1)^2}
        + (1-\theta_x^2) \frac{x (x-1)}{(u-x)^2} \right],
\label{eqPVI}
\end{eqnarray} 
in which the four $\theta_j^2$ are arbitrary complex constants,
displays the main property of this 
``\'equation diff\'erentielle curieuse'' (as Picard called it after he found it
in the particular case $\theta_j=0$, $j=\infty,0,1,x$):
its general solution $u(x)$
is singlevalued except at three points, conveniently put at $x=\infty,0,1$
so that $x$ is the crossratio $(\infty,0,1,x)$.

The second representation also originates from Picard.
It results from the invertible point transformation
$(U,X,T) \mapsto (u,x,t)$ 
[we also give here its extension to the spectral parameter $t$,
to be used later]
defined by \cite[p.~298]{Picard1889}
\begin{eqnarray}
& & {\hskip -8.0truemm}
U=\frac{1}{2 \omega}\int_{\infty}^{u} \frac{\D u}{\sqrt{u(u-1)(u-x)}},\ 
\frac{X}{\aX}=\Omega=i \pi \frac{\omega'}{\omega},\
T=\frac{1}{2 \omega}\int_{\infty}^{t} \frac{\D t}{\sqrt{t(t-1)(t-x)}},
\label{eqdefUXT}
\label{eqdefU}
\label{eqdefX}
\end{eqnarray}
(with $\aX$ some normalization constant),
whose inverse is
\begin{eqnarray}
& & {\hskip -5.0truemm}
u=\frac{\wp(2 \omega U,g_2,g_3)-e_1}{e_2-e_1},\
\sqrt{u(u-1)(u-x)}=\frac{1}{2}(e_2-e_1)^{-3/2} \wp'(2 \omega U,g_2,g_3),\
\nonumber\\ & & {\hskip -5.0truemm}
t=\frac{\wp(2 \omega T,g_2,g_3)-e_1}{e_2-e_1},\ \
\sqrt{t(t-1)(t-x)}=\frac{1}{2}(e_2-e_1)^{-3/2} \wp'(2 \omega T,g_2,g_3),
\nonumber\\ & & {\hskip -5.0truemm}
x      =\frac{              e_3-e_1}{e_2-e_1}\cdot
\label{eq-ux-Weierstrass-rep}
\label{eqdefuxt}
\end{eqnarray}
The new independent variable $X$ 
(denoted $\Omega$ by classical authors like Halphen)
is proportional to the ratio of the two half-periods $\omega, \omega'$
of the elliptic function $2 \omega U \mapsto u$.

The transformed ordinary differential equation (ODE) for $U(X)$,
a systematic computation of which is recalled in Appendix \ref{AppendixElliptic3var},
was initially written by R.~Fuchs \cite[Eq.~(8)]{FuchsP6},
then simplified by Painlev\'e \cite[p.~1117]{PaiCRAS1906}
by insertion of the prefactor $1/(2 \omega)$ in (\ref{eqdefU}),
and much later rediscovered by Manin \cite{Manin1998}
and Babich and Bordag \cite{BB1999}.

In these new coordinates $(U,X)$, $\PVI$ becomes quite simple,
\begin{eqnarray}
& &
\frac{\D^2 U}{\D X^2}=\frac{(2 \omega)^{3}}{\pi^2 \aX^2}
\sum_{j=\infty,0,1,x}\theta_j^2 \wp'(2\omega U+\omega_j,g_2,g_3),
\label{eqPVIUX}
\end{eqnarray}
in which
the summation runs over the four half-periods $\omega_j$ of $\wp$,
and $\wp'$ denotes the partial derivative of $\wp$ with respect to its first argument,
taken at point $2 \omega U+\omega_j$.
One advantage of the elliptic representation (\ref{eqPVIUX}) 
is its Hamiltonian description,
\begin{eqnarray}
& & {\hskip -8.0truemm}
H(Q,P,X)=\frac{P^2}{2}+V(Q,X),\
Q=U,\
P=\frac{\D Q}{\D X},\ 
\frac{\D^2 Q}{\D X^2}=-\frac{\partial V}{\partial Q},
%\frac{\D P}{\D X}=\frac{\D^2 Q}{\D X^2}=-\frac{\partial V}{\partial Q}\cdot
\label{eqHamiltonien-wp}
\\ & & {\hskip -8.0truemm}
V(Q,X,\lbrace\theta_j^2\rbrace)=-\frac{(2 \omega)^{2}}{\pi^2 \aX^2}
 \sum_{j=\infty,0,1,x}\theta_j^2 \wp(2 \omega Q+\omega_j,g_2,g_3).
\nonumber
\end{eqnarray}

We will respectively call $(u,x,t)$ and $(U,X,T)$ 
the \textit{rational coordinates} and \textit{elliptic coordinates} and,
depending on the context,
denote the four half-periods as
either $(\omega_\infty,\omega_0,\omega_1,\omega_x)$ (like in (\ref{eqPVIUX})),
or     $(0,\omega,\omega+\omega',\omega')$ (like in (\ref{eqdefUXT})), 
 % not $(0,\omega,\omega',\omega+\omega')$
or     $(0,\omega_1,\omega_2,\omega_3)$ (to respect the correspondence 
 $\wp(\omega_j)=e_j, j=1,2,3$ with the classical definition (\ref{eqwp})).
 
There exist various linear representations
\cite{FuchsP6,JimboMiwaII,Harnad1994,Zotov2004,NoumiYamadaLax8}
of this nonlinear ODE by a Lax pair,
and we restrict here to those which have minimal order, i.e.~two.
Their main characteristics were outlined by Poincar\'e \cite[p.~219]{Poincare1883}
in the scalar case
and by Schlesinger \cite{SchlesingerP6} in the matrix case, let us remind them.
First of all,
their singularities in the complex plane of the spectral parameter $t$
can be restricted to be only of the Fuchsian type
(``regular'' singularities).

In the scalar case, which can be defined as
\begin{eqnarray}
& & {\hskip -9.0truemm}
\partial_t^2 \psi + \frac{S}{2} \psi=0,\
\label{eqFuchsScalarODE}
%\\ & &
\partial_x \psi + C \partial_t \psi -\frac{C_t}{2} \psi = 0,\
\label{eqFuchsScalarPDE}
%\\ & & Z \equiv 
S_x + C_{ttt} + C S_t + 2 C_t S=0,
\end{eqnarray}
the first equation 
must have four Fuchsian singularities
(characterized by their crossratio $x$ since three of them can be put
at predefined locations by a homo\-graphy), 
plus,
as prescribed by Poincar\'e \cite[p.~219]{Poincare1883},
one apparent singularity, also of the Fuchsian type,
located at $t=u$.
Without such an apparent singularity,
the system would be rigid and no nonlinear ODE would come out.

In the matrix case, defined as
\begin{eqnarray}
& &
\partial_x \vectorpsi = L \vectorpsi,\
\partial_t \vectorpsi = M \vectorpsi,\
L_t-M_x+L M - M L=0,
\label{matrixLax}
\end{eqnarray}
the second order matrices $L$ and $M$ can be chosen traceless without loss of generality,
and Schlesinger \cite{SchlesingerP6} proved that 
the monodromy matrix $M$ must be the sum of four simple poles
of crossratio $x$,
and that the matrix $L$ must be the sum of a simple pole and a regular term,
i.e.~with the convention $t=\infty,0,1,x$ for the four Fuchsian singularities,
\begin{eqnarray}
& & {\hskip -14.0truemm}
L= - \frac{M_x}{t-x} + L_\infty,\
M=\frac{M_0}{t} + \frac{M_1}{t-1} + \frac{M_x}{t-x},\
M_\infty+M_0+M_1+M_x=0.
\label{eq-Matrix-Lax-pair}
\label{eqLaxPVISchlesinger}
\end{eqnarray}
In particular, no apparent singularity is required in the matrix case.

Neither Poincar\'e nor Schlesinger performed the practical computations
which they had prescribed.
This was first achieved 
in the scalar case by Richard Fuchs \cite{FuchsP6} 
(the son of the Lazarus Fuchs of Fuchsian equations and brother-in-law of Schlesinger),
and in the matrix case by Jimbo and Miwa \cite{JimboMiwaII}.
However,
while one would expect the just mentioned matrix Lax pair to be ``simpler'' than 
the scalar one because of the unnecessity for an apparent singularity,
this is not the case, as detailed in \cite{LCM2003,C2006Kyoto}.
The reason is that, in order to unveil $\PVI$ during the resolution process,
no additional assumption is required in the scalar case,
while in the matrix case one must make the following practical assumption:
in order to implement the property established by Schlesinger 
that the determinants of the four residues $M_j$ are constant
(and equivalent to the four parameters of $\PVI$),
one must assume a representation of the four second order traceless matrices $M_j$
enforcing this prescription.
\medskip

The present article, 
which is an extended version of a short note \onlinecite{Conte-Lax-PVI-CRAS},
contains three main results.
\begin{enumerate}
\item
We first show that a classical problem of geometry,
set up and solved by Pierre-Ossian Bonnet in 1867,
yields as a by-product a new, isomonodromic, very symmetric second order matrix Lax pair 
of a codimension-two $\PVI$,
which is easily extrapolated to the generic $\PVI$.
The decisive advantage conferred by its geometric origin is that no assumption is required 
concerning the four residues.

\item
The second result is a rigorous derivation of a nice property of $\PVI$,
unveiled by Suleimanov \cite{Suleimanov1994} 
and 
known as the ``quantum correspondence''.

\item
Finally, we match the completeness property of $\PVI$
(impossibilily to add complementary terms without losing the Painlev\'e property)
and the completeness property of the Gauss-Codazzi equations
(they completely describe the geometry)
by building a solution of the Gauss-Codazzi equations in terms of the full $\PVI$.

\end{enumerate}

\medskip

The paper is organized as follows.

In section \ref{section_Classical_geometry}, 
we recall the classical analytic description of surfaces,
a prerequisite to the presentation of Bonnet surfaces and their moving frame,
done in section \ref{section_Bonnet_surfaces}.

Section \ref{section_From_Bonnet_to_Lax_PVI} is the core of the paper:
we upgrade this moving frame to a second order, isomonodromic matrix Lax pair 
of $\PVI$,
and we compare this new Lax pair to the existing ones.
Next, in section \ref{sectionQuantum_correspondence},
we establish the link with the classical second order scalar Lax pair
and give a rigorous derivation of the ``quantum correspondence''.

Finally, in section \ref{sectionBonnet_plus},
by converting back the matrix Lax pair to the moving frame of some surface,
we lift Bonnet surfaces to surfaces 
which depend on two more degrees of freedom and are described by the full $\PVI$.

Most results are presented in both rational coordinates $(u,x,t)$
and elliptic coordinates $(U,X,T)$.
                   
% ============================================================================
\section{Classical geometry of surfaces}
\label{section_Classical_geometry}
% ============================================================================

As shown by Gauss in 1827, surfaces in $\mathbb{R}^3$ are characterized by 
the two ``fundamental'' quadratic forms
$<\D \bfF,\D \bfF>$, 
$-<\D \bfF,\D \bfN>$,
in which 
$\bfF(x_1,x_2)$ is the current point on the surface, 
$\D \bfF$ a vector in the tangent plane,
$   \bfN$ any unit vector normal to the tangent plane.
   In ``conformal coordinates'', these quadratic forms 
\begin{eqnarray}
& &
{\rm I}=<\D \bfF,\D \bfF>=e^\metric \D z \ \D \barz,
\label{eqform1} 
\\ & &
{\rm II}=-<\D \bfF,\D \bfN>=Q\ \D z^2 + e^\metric H \D z \ \D \barz + \barQ\ \D \barz^2,
\label{eqform2} 
\end{eqnarray}
define four fields: 
$\metric$ real, $Q$, its complex conjugate $\barQ$, $H$ real,
and the link with the two principal curvatures $1/R_1$ and $1/R_2$ is,
\begin{eqnarray}
& & {\hskip -5.0truemm}
\frac{1}{2}\left(\frac{1}{R_1}+\frac{1}{R_2}\right)=\hbox{mean curvature}=H, 
\nonumber \\
& & {\hskip -5.0truemm}
\frac{1}{R_1 R_2}=\hbox{total (or Gaussian) curvature}
= -2 e^{-\metric} \metric_{z \barz}.
\label{eqdefcurvatures}
\end{eqnarray}

If $\sigma$ denotes some moving frame defined from $\bfF$ and $\bfN$, 
the gradient of $\sigma$ defines two square matrices $\matU$, $\matV$,
\begin{eqnarray}
& & \sigma_z=\matU \sigma,\ \sigma_\barz=\matV \sigma,
\label{eqgradsigma}
\end{eqnarray}
and the zero-curvature condition
\begin{eqnarray}
& & \left\lbrack\partial_z-\matU, \partial_\barz-\matV \right\rbrack
= \matU_\barz-\matV_z+\left\lbrack\matU,\matV \right\lbrack=0,
\label{eqzerocurvature}
\end{eqnarray}
generates a set of nonlinear partial differential equations (PDEs) involving  
$\metric$, $Q$, $\barQ$, $H$.

An additional parameter $c$ can be inserted 
in these moving frame equations (\ref{eqgradsigma})
if one replaces $\mathbb{R}^3$ 
by the three-dimensional Riemannian manifold $\Rcubec$
having a constant curvature $\kappa=-c^2$.
When $\kappa$ is respectively negative, zero, positive,
this three-dimensional manifold is respectively
the hyperbolic space $\mathbb{H}^3(c)$, the Euclidean space $\mathbb{R}^3$, 
the sphere $\mathbb{S}^3(c)$ of 
radius $\kappa^{-1/2}$.
The moving frame defined by
\begin{eqnarray}
& &
\bfsigma=
\left\lbrace 
\begin{array}{ll}
\displaystyle{
{}^{\rm t} (\bfF_z,\bfF_\barz,\bfN) {\hskip 4.0truemm} (c=0),\ 
}\\ \displaystyle{
{}^{\rm t} (\bfF,\bfF_z,\bfF_\barz,\bfN)\ (c\not=0), 
}
\end{array}
\right.
\label{eqMoving-frame} 
\end{eqnarray} 
then yields matrices $\matU$, $\matV$ of respective orders three ($c=0$) and four ($c\not=0$).
Instead of them, 
it proves quite convenient to use 
the representation by second order matrices \cite{B1994,Springborn}, 
%(Bobenko and Eitner 2000) page 20 eq (2.34))
\begin{eqnarray}
& &
\left\lbrace 
\begin{array}{ll}
\displaystyle{
\matU=
\begin{pmatrix} 
 (1/4) \metric_z           & -Q e^{-\metric/2} \cr 
 (1/2) (H+c) e^{\metric/2} & -(1/4) \metric_z \cr ,\
\end{pmatrix}
}\\ \displaystyle{
\matV=\begin{pmatrix} 
-(1/4) \metric_\barz & -(1/2) (H-c) e^{\metric/2} \cr 
\barQ e^{-\metric/2} & (1/4) \metric_\barz \cr
\end{pmatrix}
}
\end{array}
\right.
\label{eqR3R4_Moving_frame_order2}
\label{eqGW-su2-R3R4}
\end{eqnarray} 

The nonlinear PDEs generated by the zero-curvature condition (\ref{eqzerocurvature})
are known as the Gauss-Codazzi equations\footnote{
The Codazzi equations were in fact first written in 1853
by Karl M.~Peterson, a Latvian student,
before Mainardi (1856) and Codazzi (1868),
see the historical notes by Phillips \cite{Phillips}.},
\begin{eqnarray}
& &
\left\lbrace 
\begin{array}{ll}
\displaystyle{
\metric_{z \barz} + \frac{1}{2} (H^2-c^2) e^\metric -2 \mod{Q}^2 e^{-\metric}=0 \hbox{ (Gauss)},
}\\ \displaystyle{
Q_\barz-\frac{1}{2} H_z e^\metric =0,\ \barQ_z-\frac{1}{2} H_\barz e^\metric =0 \hbox{ (Codazzi)}.
}
\end{array}
\right.
\label{eqGaussCodazziR3c} 
\end{eqnarray} 
This is a classical result due to Bonnet % e.g. B1999 p 9 \S 1.4
that any solution $(\metric,H,Q,\barQ)$ 
determines a unique surface up to rigid motion.
	
In addition to the classical conformal invariance,
\begin{eqnarray}
& & \forall G(z):\
(z,e^\metric,H,Q) \to \left(G(z), \mod{G'(z)}^2 e^\metric,H,{G'(z)}^2 Q \right),
\label{eqConformal}
\end{eqnarray}
and the scaling invariance, 
\begin{eqnarray}
& & \forall k:\
(z,e^\metric,H,Q,c) \to \left(z,k^2 e^\metric,k^{-1}H,k Q,k^{-1} c \right),
\label{eqScaling}
\end{eqnarray}
the system (\ref{eqGaussCodazziR3c}) possesses another invariance,
% this is: exchange of psi1 and psi2 in moving frame
which only exists under the condition $Q-\barQ=c$,
this is the involution 
\cite[Eq.~(4.4)]{B1994}
\cite[p.~77]{BE2000}
\cite[\S 3 p.~6]{Springborn} 
%[possibly in Bianchi 1903? Not seen in Lezione]
defined by the permutation of the two basis vectors of the moving frame 
(\ref{eqGW-su2-R3R4})
\begin{eqnarray}
& & 
(\metric,H,Q,\barQ) \to \left(-\metric,2 Q-c=2 \barQ+c,\frac{H+c}{2}, \frac{H-c}{2} \right).
\label{eqGC-Involution}
\end{eqnarray}

% ============================================================================
\section{Bonnet surfaces and their moving frame}
\label{section_Bonnet_surfaces}
% ============================================================================

As an application of the newly discovered Gauss-Codazzi equations,
the geometer Pierre-Ossian Bonnet set up in 1867 a natural problem 
(the Bonnet problem)
and solved it in full generality.
Since this achievement is often incompletely presented in modern articles,
we find it useful to recall 
in Appendix \ref{sectionBonnet_problem_and_solution}
the complete proof as given by Bonnet himself.

The most interesting of the five solutions to the Bonnet problem
is what is now called the \textit{Bonnet surfaces}.
Characterized in local coordinates by conditions on $\metric$, $\modQ$ and $H$
(i.e.~excluding $\arg Q$),
\begin{eqnarray}
& & {\hskip -10.0truemm}
\left\lbrace 
\begin{array}{ll}
\displaystyle{
e^\metric \modQ^{-2} H_\barz=g_1(z)    \not=0,\
e^\metric \modQ^{-2} H_z    =g_2(\barz)\not=0,
}\\ \displaystyle{
\frac{2 \D H}{g_1 \D z + g_2 \D \barz} + H^2 - c^2 \not=0,
}
\end{array}
\right.
\end{eqnarray} 
they depend,
after a conformal transformation detailed in the Appendix,
on 
one fixed parameter ($c$, at least if one considers $\Rcubec$ instead of $\mathbb{R}$)
and six arbitrary movable constants.
Their metric $\metric$ and $Q$ are given by
% Careful! No x here, only z and \barz
\begin{eqnarray}
& & {\hskip -16.0truemm}
\left\lbrace 
\begin{array}{ll}
\displaystyle{
    Q=2 \cz \coth 2 \cz(    z-    z_0) - 2 \cz  \coth 4 \cz \Re(    z-    z_0)
=\frac{\sinh 2 \cz (\barz-\barz_0)}
      {\sinh 2 \cz (    z-    z_0)}\frac{2 \cz}{\sinh 4 \cz \Re(    z-    z_0)},\
}\\ \displaystyle{
\barQ=2 \cz \coth 2 \cz(\barz-\barz_0) - 2 \cz  \coth 4 \cz \Re(    z-    z_0)
=\frac{\sinh 2 \cz (    z-    z_0)}
      {\sinh 2 \cz (\barz-\barz_0)}\frac{2 \cz}{\sinh 4 \cz \Re(    z-    z_0)},
}\\ \displaystyle{
\modQ^2=\left(\frac{2 \cz}{\sinh 4 \cz \Re(    z-    z_0)}\right)^2,
}\\ \displaystyle{
e^\metric=4 \modQ^2 \frac{\D \Re(z)}{\D H},
}
\end{array}
\right.
\label{eq-Bonnet-uHQ}
%\label{eq-Bonnet-type4-Q}
\end{eqnarray} 
in which $\cz$ is an arbitrary (possibly zero) complex constant,
and the mean curvature $H$, which only depends on $\xi=\Re(z)$,
obeys the third order ODE (\ref{eq-Bonnet-type4-ODEh}),
whose first integral (\ref{eqHazzi-ODE2h})
defines a second order second degree ODE for $H=h(\xi)$.
Despite being just a particular case of the ODE labeled (B,V) by Chazy \cite[p.~340]{ChazyThese},
this second order ODE remained unnoticed 
(even by \'Elie Cartan \cite[p.~85]{Cartan1942}) and therefore unintegrated for nearly one century,
until Bobenko and Eitner \cite{BE1998} 
expressed its general solution in terms of 
the sixth Painlev\'e equation $\PVI$ (\ref{eqPVI}).
More precisely,
the mean curvature $H$ 
is equal to
the logarithmic derivative $\HVI$ of a $\tau$-function
of $\PVI$.
Rather than the expression built by Chazy \cite[expression $t$ page 341]{ChazyThese},
it is preferable to adopt its homographic transform by Malmquist \cite{MalmquistP6},
\begin{eqnarray}
& & {\hskip -2.0 truemm}
\Dlogtau_{\rm VI,M}=\frac{x(x-1)}{4 u(u-1)(u-x)} \left(\frac{\D u}{\D x}\right)^2
\label{eqHamVICu} % curieuse notation, C ou M?
\\ & & 
+\frac{1}{4 x(x-1)}
\left[
 \theta_\infty^2 \left(-u+\frac{1}{2}\right) 
+\theta_0^2            \left(-\frac{x}{u}+\frac{1}{2}\right)
\right. \nonumber\\ & & \phantom{1234567890}\left.
+ \theta_1^2            \left(\frac{x-1}{u-1} -\frac{1}{2}\right)
+(\theta_x-1)^2         \left(-\frac{x(x-1)}{u-x} -x +\frac{1}{2}\right)
\right],
\nonumber
\end{eqnarray}
for two (equivalent) reasons:
 (i) absence of a first degree term $\D u / \D x$,
(ii) choice of $\theta_x$ to break the parity invariance of (\ref{eqPVI}) in the $\theta_j$'s 
(Chazy chose $\theta_\infty$).
Bonnet surfaces are then analytically represented as
\begin{eqnarray}
& & {\hskip -15.0 truemm}
x=\frac{1}{1-e^{4 \cz (z+\barz)}},\ 
H = 8 \cz Y,\ Y= x(x-1) \Dlogtau_{\rm VI,M},\ 
\end{eqnarray}
with however three constraints 
among the four monodromy exponents $\theta_j$,
\begin{eqnarray}
& &
\theta_\infty=0,\ 
c= \cz (\theta_1^2-\theta_0^2),\
\theta_x^2=1.
\label{eqBonnetContraintes-thetaj}
\end{eqnarray} 
The only movable singularities of $H=h(\xi)$ are a unique movable simple pole.

The moving frame (\ref{eqR3R4_Moving_frame_order2}) of these Bonnet surfaces
\begin{eqnarray}
& & {\hskip -15.0 truemm}
\left\lbrace 
\begin{array}{ll}
\displaystyle{
\matU \D z + \matV \D \barz
=x (x-1) \frac{Y''}{Y'}(\cz \D \barz-\cz \D z)
 \begin{pmatrix} 1 & 0 \cr 0 & -1 \end{pmatrix}
}\\ \displaystyle{
+ \sqrt{Y'} 
\begin{pmatrix} 
0 & -S_1 \D     z -\frac{\displaystyle{Y -(\theta_0^2-\theta_1^2)/8}}{\displaystyle{Y'}} 4 \cz \D \barz \cr 
              S_2 \D \barz +\frac{\displaystyle{Y +(\theta_0^2-\theta_1^2)/8}}{\displaystyle{Y'}} 4 \cz \D z& 0 \cr
\end{pmatrix},
}\\ \displaystyle{
S_1 =\frac{2 \cz}{\sinh(2 \cz (z+\barz))}\frac{\sinh(2 \cz \barz)}{\sinh(2 \cz z)},\
S_2 =\frac{2 \cz}{\sinh(2 \cz (z+\barz))}\frac{\sinh(2 \cz     z)}{\sinh(2 \cz \barz)},
}
\end{array}
\right.
\label{eqdpsiBonshort}
\end{eqnarray}
then defines a linear representation of the variable $\HVI$, 
however with several undesired features:
\begin{description}
	\item[] -- lack of a spectral parameter,
		
	\item[] -- nonrational dependence on $z$, $\barz$, $Y'$, 
			
	\item[] -- restriction to $\theta_\infty=0,\ \theta_x^2=1$. 
	                         
\end{description}

\textit{Remark}.
Under the involution (\ref{eqGC-Involution}),
Bonnet surfaces are mapped to surfaces with a harmonic inverse mean curvature 
\cite[Prop.~4.7.1 page 77]{BE2000},
which also integrate with $\PVI$ \cite{BEK1997} and whose moving frame is
\begin{eqnarray}
& & {\hskip -15.0 truemm}
\matU \D z + \matV \D \barz
=\left[-X (X-1) \frac{Y''}{Y'}(\cz \D \barz-\cz \D z) -\frac{1}{2} \D \log \frac{\sinh 2 \cz z}{\sinh 2 \cz \barz}\right] 
\begin{pmatrix} 1 & 0 \cr 0 & -1\end{pmatrix}
\nonumber \\ & & {\hskip -15.0 truemm}
+ \sqrtYone 
\begin{pmatrix} 
0 & -4\cz\D \barz-S_1 \frac{\displaystyle{Y + (\theta_0^2-\theta_1^2)/8}}{\displaystyle{X(X-1) Y'}}\D z  \cr
              4\cz\D     z+S_2 \frac{\displaystyle{Y - (\theta_0^2-\theta_1^2)/8}}{\displaystyle{X(X-1) Y'}}\D \barz  & 0 
\end{pmatrix}.
\label{eqdpsiBEKshort}
\end{eqnarray}
Since the transition matrix
\begin{eqnarray}
& & 
P=
\begin{pmatrix} 0 & g \cr -1/g & 0 \cr \end{pmatrix},\
g=\left(\frac{\sinh(2 \cz \barz)}{\sinh(2 \cz z)}\right)^{1/2}, 
\end{eqnarray}
maps (\ref{eqdpsiBEKshort}) to (\ref{eqdpsiBonshort}),
it is sufficient to consider the moving frame of Bonnet surfaces. 

% ============================================================================
\section{From the moving frame of Bonnet to a Lax pair of $\PVI$}
\label{section_From_Bonnet_to_Lax_PVI}
% ============================================================================

	Let us convert the moving frame to
a second order, isomonodromic matrix Lax pair for the generic $\PVI$ for $u(x)$, 
i.e.~with the following properties,

\begin{description}

	\item[] -- dependence on an arbitrary parameter $t$
	(the spectral parameter),
	
	\item[] -- rational dependence on $t$, the independent variable $x$
	and the dependent variables $u(x)$, $u'(x)$ of $\PVI$,
	
	\item[] -- absence of any restriction on the four $\theta_j$'s.
			
\end{description}

The successive steps are:
\begin{enumerate}
% ================================================================== STEP 1
	\item Introduction of a spectral parameter $t$
and creation of a rational dependence on $x$ and $t$.
This is not achieved by a conformal transformation (\ref{eqConformal})
like for constant mean curvature surfaces, see Eq.~(\ref{eqBonnet-type1SG}), 
but \textit{via} a change of variables $(z,\barz) \to (x,t)$.
Indeed, as shown in \cite{BE1998}, % (37)
choosing for $x$ and $t$ any homographic transform of, respectively,
$e^{4 \cz (z+\barz)}$ and $e^{4 \cz z}$
creates four poles in the monodromy matrix.
For instance, the choice
\begin{eqnarray}
& & {\hskip -15.0 truemm}
x=\frac{1}{1-e^{4 \cz (z+\barz)}},\ 
t=\frac{1}{1-e^{4 \cz z}},\ 
\end{eqnarray}
creates the set of poles $t=\infty,0,1,x$, 
\begin{eqnarray}
& & {\hskip -15.0 truemm}
\left\lbrace
\begin{array}{ll} 
\displaystyle{
4 \cz \D z= \frac{\D t}{t(t-1)},\
4 \cz \D \barz=\frac{\D x}{x(x-1)}-\frac{\D t}{t(t-1)},\
}\\ \displaystyle{
S_1 \D z=-\frac{t-x}{t(t-1)}\D t,\
S_2 \D \barz=-\frac{\D x}{t-x}+ \frac{x(x-1)}{t(t-1)(t-x)}\D t,
}
\end{array}
\right.
\label{eqBEK-choice-t}
\end{eqnarray}
and therefore changes the moving frame (\ref{eqdpsiBonshort}) 
to an isomonodromic Lax pair (\ref{eq-Matrix-Lax-pair})
for an incomplete $\Dlogtau_{\rm VI,M}$ 
with an algebraic dependence on $Y'$,
\begin{eqnarray}
& & 
(z,\barz) \to (x,t),\
\matU \D z + \matV \D \barz=L \D x + M \D t,
\\ & &
\det M_\infty =\det M_x=0,\ -4\det M_0=\theta_0^2,\ -4 \det M_1=\theta_1^2.
\end{eqnarray}
For reference, the $(x,t)$ representation of the Bonnet surfaces is
\begin{eqnarray}
& & {\hskip -16.0truemm}
\left\lbrace 
\begin{array}{ll}
\displaystyle{
H=\frac{c}{\kappa} 8 \cz x(x-1) \Dlogtau_{\rm VI,M},\
e^{-\metric}=\frac{c^2}{\kappa^2}\frac{\D}{\D x} [x(x-1)\Dlogtau_{\rm VI,M}],\
}\\ \displaystyle{
    Q=-\frac{\kappa}{c} 4 \cz (t-x),\
\barQ=-\frac{\kappa}{c} 4 \cz \frac{x(x-1)}{t-x},\
\modQ^2=16 \frac{\kappa^2}{c^2} \cz^2 x(x-1),
}\\ \displaystyle{
\theta_\infty=0,\ 
\theta_x^2=1,\
\kappa= \cz (\theta_1^2-\theta_0^2),\
}
\end{array}
\right.
\label{eq-Bonnet-uHQ-xt}
\end{eqnarray} 
values in which $c$ has been made arbitrary by the scaling invariance
(\ref{eqScaling}).

% ================================================================== STEP 2

	\item Switch from the $\Dlogtau_{\rm VI,M}$ field $Y$ to the $\PVI$ field $u$.
	This is done
\textit{via} the birational transformation between $\PVI$
and its Hamiltonian $\Dlogtau_{\rm VI,M}$
defined by the relations \cite[p.~341]{ChazyThese}
\begin{eqnarray}
& &
Y=x(x-1)\Dlogtau_{\rm VI,M}, 
\label{eqHPVInorm}
\end{eqnarray}
and \cite[Table R]{Okamoto1980-II}, 
\begin{eqnarray}
& & {\hskip -19.0 truemm}
u=x+\frac{\Theta_x x(x-1) Y''
 -\left[Y +\displaystyle{\frac{\theta_\infty^2+3\Theta_x^2}{8}}(2 x-1)-\frac{\theta_1^2-\theta_0^2}{8}\right]
  \left[Y'+\displaystyle{\frac{\theta_\infty^2+\Theta_x^2}{4}}\right]
}{\left(Y'+\displaystyle{\frac{(\theta_\infty+\Theta_x)^2}{4}}\right)
  \left(Y'+\displaystyle{\frac{(\theta_\infty-\Theta_x)^2}{4}}\right)}
\nonumber\\ & & \phantom{1234567} 
+\frac{\displaystyle{\frac{\Theta_x^2}{2}} 
  \left[Y +\displaystyle{\frac{3\theta_\infty^2+\Theta_x^2}{8}}(2 x-1)+\displaystyle{\frac{\theta_1^2-\theta_0^2}{8}}\right]
}{\left(Y'+\displaystyle{\frac{(\theta_\infty+\Theta_x)^2}{4}}\right)
  \left(Y'+\displaystyle{\frac{(\theta_\infty-\Theta_x)^2}{4}}\right)}\ccomma
\label{equtoY}
\end{eqnarray}
in which $\Theta_x$ denotes the shifted exponent
\begin{eqnarray}
& &
\Theta_x=\theta_x-1.
\label{eqdefThetax}
\end{eqnarray}
While the Hamiltonian (\ref{eqHamVICu}) breaks the parity of only one $\theta_j$,
the birational transformation between $\PVI$ and $\HVI$ breaks the parity of two $\theta_j$'s. 

For the Bonnet constraints $\theta_\infty=0$, $\theta_x^2=1$ (\ref{eqBonnetContraintes-thetaj}),
these two relations %(\ref{eqHPVInorm}), (\ref{equtoY}) 
simplify to
\begin{eqnarray}
& & {\hskip -15.0 truemm}
Y=x(x-1) \Dlogtau_{\rm VI,M},\ 
\frac{8 Y' }{8 Y +\theta_0^2-\theta_1^2}=-\frac{1}{u-x},\   
\frac{Y''}{Y'}= -\frac{u'}{u-x}\ccomma 
\end{eqnarray} 
and now change the moving frame
to an algebraic isomonodromic Lax pair for a codimension-two $\PVI$,
\begin{eqnarray}
& & {\hskip -15.0 truemm}
\left\lbrace 
\begin{array}{ll}
\displaystyle{
L \D x + M \D t
=-x (x-1) \frac{u'}{u-x}(\cz \D \barz-\cz \D z)
\begin{pmatrix} 1 & 0 \cr 0 & -1 \cr \end{pmatrix}
}\\ \displaystyle{
 +\sqrt{-\frac{\fys}{u-x}} 
\begin{pmatrix} 
0 & -S_1 \D     z +\frac{\displaystyle{(u-x)Y}}{\displaystyle{\fys}} 4 \cz \D \barz \cr 
              S_2 \D \barz -\frac{\displaystyle{(u-x)Y}}{\displaystyle{\fys}} 4 \cz \D z& 0 \cr
\end{pmatrix},
}\\ \displaystyle{
\fys=x(x-1) \Dlogtau_{\rm VI,M} +\frac{\theta_0^2-\theta_1^2}{8},
}
\end{array}
\right.
\label{eqdpsiBonstep1}
\end{eqnarray}
in which $\D z$ and $\D \barz$ are assumed replaced by their values
(\ref{eqBEK-choice-t}).

% ================================================================== STEP 3
	
	\item Removal of the algebraic dependence on $Y'$
by a change of basis vectors defined by the transition matrix
\begin{eqnarray}
& & {\hskip -15.0 truemm}
P_1=\diag({Y'}^{1/4},{Y'}^{-1/4}),\ 
Y'= -\frac{\fys}{u-x}\cdot
\label{eqTransitionP1}
\end{eqnarray}
The algebraic Lax pair (\ref{eqdpsiBonstep1}) becomes rational,
and its five terms as defined by 
(\ref{eq-Matrix-Lax-pair}) evaluate to
\begin{eqnarray}
& & {\hskip -15.0 truemm}
\left\lbrace
\begin{array}{ll} 
\displaystyle{
\theta_\infty=0,\ \Theta_x=0,\ 
\fys=x(x-1) \Dlogtau_{\rm VI} +\frac{\theta_0^2-\theta_1^2}{8},\  
}\\ \displaystyle{
M_\infty=
\begin{pmatrix} 0 & -1 \cr 0 & 0 \cr\end{pmatrix},\
M_x     =
\begin{pmatrix} 0 & 0 \cr -\displaystyle{\frac{\fys}{u-x}} & 0 \cr
\end{pmatrix},\
}\\ \displaystyle{
M_0=\frac{1}{2}
\begin{pmatrix} 
-x (x-1) \displaystyle{\frac{u'}{u-x}} & 
 2 u \cr 
  \displaystyle{\frac{\theta_0^2-\theta_1^2}{2}} -2 \fys\ 
  -2 \displaystyle{\frac{x-1}{u-x}}\fys &
                x (x-1) \displaystyle{\frac{u'}{u-x}}\cr
\end{pmatrix},\
}\\ \displaystyle{
}\\ \displaystyle{
M_1=\frac{1}{2}
\begin{pmatrix} 
x (x-1) \displaystyle{\frac{u'}{u-x}} & 
-2 (u-1) \cr 
 -\displaystyle{\frac{\theta_0^2-\theta_1^2}{2}} +2 \fys\ 
  +2 \displaystyle{\frac{x}{u-x}}\fys &
               -x (x-1) \displaystyle{\frac{u'}{u-x}}\cr
\end{pmatrix},\
}\\ \displaystyle{
L_\infty=-\frac{u-x}{x(x-1)} M_\infty\cdot
}
\end{array}
\right.
\label{eqLax-PVI-codim2-rational}
%\label{eqdpsiBonstep2}
\end{eqnarray}

% ================================================================== STEP 4
		
	\item (Last step)
Extrapolation to arbitrary values of the $\theta_j$'s.	
It is sufficient to notice that all residues 
in (\ref{eqLax-PVI-codim2-rational})
are polynomials of $u'$ of degree at most two.
By only requiring the conservation of such a dependence
and enforcing $M_\infty=$ constant,
one immediately removes all the constraints on the $\theta_j$'s. 
Each term $u'$ of (\ref{eqLax-PVI-codim2-rational})
essentially extrapolates to the left-hand-side of a Riccati 
equation which defines hypergeometric solutions of $\PVI$.

\end{enumerate}

The final Lax pair can be defined in two canonical forms.

%\vfill\eject

% ============================================================================
\subsection{First canonical form of the matrix Lax pair}
% ============================================================================

This first canonical form is valid for any value of the $\theta_j$'s, 
\begin{eqnarray}
& & %{\hskip -35.0 truemm}
\left\lbrace
\begin{array}{ll} 
\displaystyle{
L=-\frac{M_x}{t-x}-\frac{u-x}{x(x-1)} M_\infty\ccomma\
M=\frac{M_0}{t}+\frac{M_1}{t-1}+\frac{M_x}{t-x}\ccomma\
}\\ \displaystyle{
M_\infty+M_0+M_1+M_x=0,
}\\ \displaystyle{
M_\infty=\frac{1}{4}
\begin{pmatrix} 2 \auu & -4 \cr \auu^2-\theta_\infty^2 & -2 \auu\cr
\end{pmatrix},
}\\ \displaystyle{
M_0=-\frac{1}{2(u-x)}
\begin{pmatrix} 
\Mzero & -2 u(u-x) \cr 
 \displaystyle{\frac{\Mzero^2-\theta_0^2(u-x)^2}{2 u (u-x)}} & -\Mzero \cr
\end{pmatrix},\
}\\ \displaystyle{
}\\ \displaystyle{
M_1= \frac{1}{2(u-x)}
\begin{pmatrix} 
\Mone & -2 (u-1)(u-x) \cr 
 \displaystyle{\frac{\Mone^2-\theta_1^2(u-x)^2}{2 (u-1) (u-x)}} & -\Mone \cr
\end{pmatrix},\
}\\ \displaystyle{
}\\ \displaystyle{
M_x     =\frac{1}{2}
\begin{pmatrix} -\Theta_x & 0 \cr 2 M_{x,21} & \Theta_x \cr \end{pmatrix},\
}\\ \displaystyle{
M_{x,21}=-\frac{
\A^2-(u-x)^2 ((\theta_\infty^2+\Theta_x^2-2 \auu\Theta_x) u(u-1)-\theta_0^2 (u-1)+\theta_1^2 u)}
{4 u(u-1)(u-x)^2},
}\\ \displaystyle{
\A=x(x-1)u'+\Theta_x u(u-1),\ \Theta_x^2=(\theta_x-1)^2,\ 
}\\ \displaystyle{
\Mzero=\A-(\Theta_x-\auu)u(u-x),
}\\ \displaystyle{
\Mone=\A-(\Theta_x-\auu)(u-1)(u-x),
}\\ \displaystyle{	
-4 \det M_j=\theta_j^2,\ j=\infty,0,1; -4 \det M_x=\Theta_x^2,														
}
\end{array}
\right.
\label{eqLax-PVI-codim0-unbalanced-rational}
\label{eqLax-PVI-codim0-unbalanced-holomorphic}
\end{eqnarray}
in which $\auu$ is an arbitrary constant,
which can be set to any convenient value, such as $0$, $\Theta_x$ or $\pm\theta_\infty$,
by action of a constant transition matrix.

The main property of this Lax pair is ho have \textit{exactly}
the same dependence on all variables as $\Dlogtau_{\rm VI,M}$, Eq.~(\ref{eqHamVICu}):
second degree polynomial in $u'$,
meromorphic dependence on $u$ (the only poles being those of $\PVI$),
meromorphic dependence on $x$ (same),
affine dependence on three $\theta_j^2$'s
and break of the affine dependence on $\theta_x^2$
(enjoyed by $\PVI$ but not by $\Dlogtau_{\rm VI,M}$).
Another property if that the four residues are always nonzero;
although not required in the matrix case (see section \ref{sectionJM} hereafter),
such a property is an essential requirement in the scalar case:
three Fuchsian singularities are insufficient to generate $\PVI$.

The structure of $\PVI$ ($u''$ a second degree polynomial of $u'$)
makes it possible to find an even simpler Lax pair,
whose matrices $L$, $M$ would be first degree polynomials of $u'$,
so that the zero-curvature condition does not generate powers of $u'$
higher than ${u'}^2$.
Such a Lax pair does not exist 
in the class (\ref{eq-Matrix-Lax-pair}),
but it does exist outside this class, see (\ref{eqLM1}) hereafter.

% ============================================================================
\subsection{Second canonical form of the matrix Lax pair}
% ============================================================================

About the matrix Lax pair (\ref{eq-Matrix-Lax-pair}),
Schlesinger \cite[p.~105]{SchlesingerP6} proved two results,
which are indeed obeyed by (\ref{eqLax-PVI-codim0-unbalanced-holomorphic}):
 (i) the residue $M_\infty$ must be a constant,
(ii) the regular term $L_\infty$ must be a scalar multiple of $M_\infty$.
He also proved that, if $M_\infty$ is invertible, 
there exists a change of basis allowing one to cancel the term $L_\infty$
in (\ref{eq-Matrix-Lax-pair})
and therefore to uniquely define the Lax pair.

If $\theta_\infty$ vanishes, $M_\infty$ is a Jordan matrix and the pair
(\ref{eqLax-PVI-codim0-unbalanced-holomorphic}) is final.
If $\theta_\infty$ is nonzero,
the transition matrix $P_2 P_3$,
\begin{eqnarray}
& & {\hskip -10.0 truemm}
P_2=
\begin{pmatrix} 2 & 2 \cr \auu-\theta_\infty & \auu+\theta_\infty \cr \end{pmatrix},
P_3=
\begin{pmatrix} g^{-1/2} & 0 \cr 0 & g^{1/2} \cr \end{pmatrix},\
\frac{g'}{g}=\theta_\infty\frac{u-x}{x(x-1)}\ccomma
\label{eqpsiq-to-psim}
\end{eqnarray}
yields the second canonical form,
%\vfill\eject
\begin{eqnarray}
& & %{\hskip -20.0 truemm}
\left\lbrace
\begin{array}{ll} 
\displaystyle{
\theta_\infty\not=0:\ 
L = -\frac{M_x}{t-x},\ 
M=\frac{M_0}{t}+\frac{M_1}{t-1}+\frac{M_x}{t-x},\
\Theta_x^2=(\theta_x-1)^2,\ 
}\\ \displaystyle{
M_\infty+M_0+M_1+M_x=0,
}\\ \displaystyle{
M_\infty =\frac{1}{2} 
\begin{pmatrix} \theta_\infty & 0\cr 0 & -\theta_\infty \cr
\end{pmatrix},\
}\\ \displaystyle{
M_{0,11}=\ \ \frac{u-1}{N}\left[\left(\A- \Theta_x                u(u-x)\right)^2-(u-x)^2(\theta_0^2+\theta_\infty^2 u^2)\right],\
}\\ \displaystyle{
M_{0,12}=\ \ \frac{u-1}{N}\left[\left(\A-(\Theta_x+\theta_\infty) u(u-x)\right)^2-(u-x)^2 \theta_0^2\right] g,\
}\\ \displaystyle{
M_{0,21}=-   \frac{u-1}{N}\left[\left(\A-(\Theta_x-\theta_\infty) u(u-x)\right)^2-(u-x)^2 \theta_0^2\right] g^{-1},\
}\\ \displaystyle{
M_{1,11}=-   \frac{u}  {N}\left[\left(\A- \Theta_x            (u-1)(u-x)\right)^2-(u-x)^2(\theta_1^2+\theta_\infty^2 (u-1)^2)\right],\
}\\ \displaystyle{
M_{1,12}=-   \frac{u}  {N}\left[\left(\A-(\Theta_x+\theta_\infty)(u-1)(u-x)\right)^2-(u-x)^2 \theta_1^2\right] g,\
}\\ \displaystyle{
M_{1,21}=\ \ \frac{u}  {N}\left[\left(\A-(\Theta_x-\theta_\infty)(u-1)(u-x)\right)^2-(u-x)^2 \theta_1^2\right] g^{-1},\
}\\ \displaystyle{
M_{x,11}=\ \ \frac{1}  {N}\left[      \A                                          ^2-(u-x)^2
  \left[(\theta_\infty^2+\Theta_x^2)u(u-1)-\theta_0^2 (u-1) + \theta_1^2 u)\right]\right],\
}\\ \displaystyle{
M_{x,12}=\ \ \frac{1}  {N}\left[      \A                                          ^2-(u-x)^2 
  \left[(\Theta_x+\theta_\infty)^2 )u(u-1)-\theta_0^2 (u-1) + \theta_1^2 u)\right]\right] g,\
}\\ \displaystyle{
M_{x,21}=-   \frac{1}  {N}\left[      \A                                          ^2-(u-x)^2 
  \left[(\Theta_x-\theta_\infty)^2 )u(u-1)-\theta_0^2 (u-1) + \theta_1^2 u)\right]\right] g^{-1},\
}\\ \displaystyle{
-4 \det M_j=\theta_j^2,\ j=\infty,0,1; -4 \det M_x=\Theta_x^2,
}
\end{array}
\right.
\label{eqLax-PVI-codim0-balanced-meromorphic}
\end{eqnarray}
with the notation
\begin{eqnarray}
& & {\hskip -10.0 truemm}
\frac{g'}{g}= \theta_\infty\frac{u-x}{x(x-1)},\
\A=x(x-1) u' + \Theta_x u(u-1),\
N=4 \theta_\infty u(u-1)(u-x)^2.
\end{eqnarray}

This Lax pair displays a nice symmetry with respect to the diagonal,
\begin{eqnarray}
& & 
M_{12}(\theta_\infty)=M_{21}(-\theta_\infty).
\end{eqnarray}
							
This result puts an end to our previous attempts \cite{LCM2003,C2006Kyoto}.

% ============================================================================
\subsection{Comparison with existing matrix Lax pairs}
% ============================================================================

Among the two other second order matrix Lax pairs of $\PVI$ we are aware of,
one 
\cite[Eq.~(C.47)]{JimboMiwaII} \cite{Cargese1996Mahoux}
has four Fuchsian singularities in the complex plane of 
its spectral parameter $t$, 
like
(\ref{eqLax-PVI-codim0-unbalanced-holomorphic}) or
(\ref{eqLax-PVI-codim0-balanced-meromorphic}),
while the other \cite{Zotov2004} 
has four Fuchsian singularities on the torus
and depends on its spectral parameter $T$ through
the function $\sigma$ of Weierstrass. 

% ============================================================================
\subsubsection{Matrix Lax pair of Jimbo and Miwa}
\label{sectionJM}
% ============================================================================

Let us start with the rational one,
which has the structure (\ref{eq-Matrix-Lax-pair})
(four Fuchsian singularities in the complex $t$ plane).
Any isomonodromic deformation of the Fuchsian system
\begin{eqnarray}
& & {\hskip -15.0 truemm}
\frac{\partial}{\partial_t} \psi=
\left(\frac{M_0}{t}+\frac{M_1}{t-1}+\frac{M_x}{t-x}\right) \psi,\
M_\infty=-M_0-M_1-M_x,
\label{eq-isomonodromic-deformation}
\end{eqnarray}
has to overcome a technical difficulty,
already mentioned in the introduction,
which consists in finding a ``nice'' representation of the property
$\det M_j =$ constant.
The representation chosen by Jimbo and Miwa
\cite[Eq.~(C.47)]{JimboMiwaII}
can be made traceless \cite[Eq.~(3.6)]{Cargese1996Mahoux},
this is
\begin{eqnarray}
& & {\hskip -15.0 truemm}
M_\infty=\frac{1}{2}
\begin{pmatrix} \theta_\infty & 0 \cr 0 & -\theta_\infty\cr
\end{pmatrix},\
M_j=\frac{1}{2}
\begin{pmatrix} z_j & (\theta_j-z_j) u_j \cr (\theta_j+z_j)/u_j& -z_j\cr
\end{pmatrix},
j=0,1,x. 
\label{eq-residues-assumption-JM}
\end{eqnarray}
It defines four functions $z_0,z_1,u_0,u_1$
of three variables $x,u,u'$ to be determined 
by the zero-curvature condition,
and this results in quite intricate expressions
for $L_{ij}, M_{ij}$ 
as detailed in \cite[Table 1]{LCM2003} and in \cite[p.~211]{CMBook}.
For $\theta_\infty=0$,
the Lax pair still exists although one of the four residues vanishes.
	
The decisive advantage of the geometric origin of the linear representation
is to avoid this difficulty,
and the structure of the residues of (\ref{eqLax-PVI-codim0-balanced-meromorphic})
is \textit{a posteriori}
\begin{eqnarray}
& & {\hskip -15.0 truemm}
M_\infty=\frac{1}{2}
\begin{pmatrix} \theta_\infty & 0 \cr 0 & -\theta_\infty\cr
\end{pmatrix},\
M_j=\frac{f_j}{\theta_\infty} 
\begin{pmatrix} P_{j,11} & P_{j,12} g\cr -P_{j,21} g^{-1}& -P_{j,11}\cr
\end{pmatrix},\
j=0,1,x, 
\label{eq-residues-assumption-poly2}
\end{eqnarray}
involving 
two rational functions $f_j(x,u)$ and 
six monic second degree polynomials of $x(x-1)u'$ whose coefficients are polynomial in $(x,u)$.

% ============================================================================
\subsubsection{Matrix Lax pair affine in $\theta_j$}
% ============================================================================

Let us now turn to the Lax pair 
defined in elliptic coordinates $(U,X,T)$ \cite{Zotov2004}.
It is affine in $\theta_j$
and 
it only involves one dimensionless function $\varphi$ \cite[Eq.~(A.10)]{Zotov2004}
which mainly depends on the two dimensionless variables $U,T$
and also on one of the four half-periods $\omega_j$,
\begin{eqnarray}
& & {\hskip -10.0truemm}   
%function PhiZ(U,T) (2*omega)*#e**(-2*eta*(2*omega)*U*T)
%     *wsigmag(2*omega*U+2*omega*T,wg2,wg3)
%     /wsigmag(2*omega*U          ,wg2,wg3)
%     /wsigmag(          2*omega*T,wg2,wg3);
%		
%phij=PhiZ(U+omegajr         ,T)*#e**(T*nj);                            
\varphi(U+\omega_j/(2 \omega),T)
= 2 \omega \frac{\sigma(2 \omega U+\omega_j+ 2 \omega T,g_2,g_3)}
                {\sigma(2 \omega U+\omega_j            ,g_2,g_3)
			           \sigma(                     2 \omega T,g_2,g_3)}
e^{\displaystyle{-2\eta(2 \omega U+\omega_j) T + n_j T}},\
\label{eqphialpha}
\end{eqnarray}
in which $n_j$ is an integer multiple of $i \pi$ characterized by the property
\begin{eqnarray}
& & 
% Zotov (A.4), (A.5)
% AMP "Differential of omegaj/(2 omega)"
% Halphen I p 309 eq (40); I p 260
% domegajr=nj*dOmega/(2*impi*ipi)
n_j=2 (\eta \omega_j - \eta_j \omega),\
\D \frac{\omega_j}{2 \omega} = -\frac{n_j}{ 2 \pi^2} \frac{\D X}{\aX},
\label{eqLaxZotov-afterA4} 
\end{eqnarray}
(with the classical notation $\eta_j=\zeta(\omega_j),\eta=\zeta(\omega)$).

Such a function $\varphi$ is classically called \cite{HalphenTraite}
an elliptic function of the second kind of $U$ (resp.~$T$)
(i.e.~not doubly periodic, but multiplied by the exponential 
of an affine function of $U$ (resp.~$T$)
under addition of one period).
%very closely related to the \textit{\'el\'ement simple} 
%introduced by Halphen \cite[t.~II p.~506]{HalphenTraite}
%to integrate the Lam\'e equation.
%\hfill\break\noindent [Halphen I p 196 formule (49)]

Denoting $\varphi'$ the derivative of $\varphi$ with respect to its first argument,
this Lax pair is
\begin{eqnarray}
& & {\hskip -10.0truemm}
\D \Psi=\mathcal{L} \Psi \D X + \mathcal{M} \Psi \D T,\ 
\label{eqdefdpsi}
\\ & & {\hskip -10.0truemm}
\mathcal{L}= \frac{1}{(2\pi)^2}\sum_{j=\infty,0,1,x}\theta_j
\begin{pmatrix} 
0 & \varphi'( U+\omega_j/(2 \omega),T) \cr 
             \varphi'(-U+\omega_j/(2 \omega),T) & 0 \cr
\end{pmatrix},
\label{eqLaxZotov}
\\ & & {\hskip -10.0truemm}
\mathcal{M}=-\frac{1}{2} \sum_{j=\infty,0,1,x}\theta_j
\begin{pmatrix} 
0 & \varphi ( U+\omega_j/(2 \omega),T) \cr 
             \varphi (-U+\omega_j/(2 \omega),T) & 0 \cr
\end{pmatrix}
\nonumber\\ & & {\hskip -10.0truemm} \phantom{1234}
+\pi^2 
\begin{pmatrix} 1 & 0 \cr 0 & -1 \cr\end{pmatrix}
\frac{\D U}{\D X}\cdot
\nonumber
\end{eqnarray}

Once converted to rational coordinates
(see Appendix \ref{AppendixElliptic3var}),
this Lax pair still depends on $W(u,x)$ and $W(t,x)$,
with $W$ defined in (\ref{eqWdef}).
Fortunately, there exists a transition matrix
\begin{eqnarray}
& & {\hskip -11.0truemm}
P=\diag(e^{-F(u,x,t)/2},e^{F(u,x,t)/2}),
\label{eqPdiag}
\\ & & {\hskip -11.0truemm}
e^{-2 F(u,x,t)}=\frac{\varphi (U,T)}{\varphi (-U,T)}
=\frac{\sigma(2 \omega(T+U),g_2,g_3)}{\sigma(2 \omega(T-U),g_2,g_3)} e^{-8 \eta\omega U T},
\end{eqnarray}
able to eliminate all $W$ terms.
The resulting Lax pair
\begin{eqnarray}
& & {\hskip -11.0truemm}
\D \Psi_1=\mathcal{L}_1 \Psi_1 \D x +\mathcal{M}_1\Psi_1 \D t,\
\Psi=P \Psi_1,\ 
\end{eqnarray}
has all its elements 
algebraic in $u,x,t$,
affine in $u',\theta_j$,
and it displays the same remarkable symmetry between $u$ and $t$
as the scalar Lax pair of Fuchs,
\begin{eqnarray}
& & {\hskip -10.0truemm}
%matlalg=(
%  matrix((2,2),1,0,0,-1)
%   *(-x*xm1(x)*'(u,1)(x)*t*tm1(t)/tmus(us,t))
% +matrix((2,2),0,c1,c2,0)*ti/sqrt(tmus(us,t))
%   *(-t*tm1(t)*sqrt(fprod(us,x)))
% +matrix((2,2),0,c1,-c2,0)*ti/sqrt(tmus(us,t))
%    *us*(us-1)*sqrt(fprod(t,x))
%        )*(us-x)/(4*x*xm1(x)*sqrt(fprod(us,x))*sqrt(fprod(t ,x)));
%
\mathcal{L}_1=\frac{1}{4 x(x-1)}
\Big[ 
   -\frac{x(x-1) t(t-1) (u-x) u'}
		     {\sqrt{t(t-1)(t-x)}\sqrt{u(u-1)(u-x)}(t-u)}                            
				\begin{pmatrix} 1 & 0 \cr  0 & -1 \cr \end{pmatrix}
  \nonumber \\ & & 
	 -\theta_\infty\frac{t(t-1)(u-x)}{\sqrt{t(t-1)(t-x)}\sqrt{t-u}}
		\begin{pmatrix} 0 & 1 \cr  1 &  0 \cr \end{pmatrix}
	 +\theta_\infty\frac{\sqrt{u(u-1)(u-x)}}{\sqrt{t-u}}
		\begin{pmatrix} 0 & 1 \cr -1 &  0 \cr\end{pmatrix}
	\nonumber \\ & & 
		+ \sum_{j=1}^3 \theta_j 
		\left(\frac{t(t-1)}{\sqrt{t(t-1)(t-x)}}-4 A_{j,+}\right)\sqrt{t-u}\sqrt{B_{j,+}}
		\begin{pmatrix} 0 & 1 \cr 0 &  0 \cr\end{pmatrix}
	\nonumber \\ & & 
		+ \sum_{j=1}^3 \theta_j 
		\left(\frac{t(t-1)}{\sqrt{t(t-1)(t-x)}}-4 A_{j,-}\right)\sqrt{t-u}\sqrt{B_{j,-}}		
		\begin{pmatrix} 0 & 0 \cr 1 &  0 \cr\end{pmatrix}
				\Big],
\nonumber \\ & & {\hskip -5.0truemm}
\nonumber \\ & & {\hskip -10.0truemm}
%matmalg=(
%  matrix((2,2),1,0,0,-1)
%   *((t-x)*us*(us-1)/tmus(us , t)-x*xm1(x)*'(u,1)(x))
% +matrix((2,2),0,c1,c2,0)*ti*(-sqrt(fprod(us,x))*sqrt(tmus(us,t)))
%        )/(4*sqrt(fprod(us,x))*sqrt(fprod(t ,x)));
%
\mathcal{M}_1=
   \frac{1}{4 \sqrt{t(t-1)(t-x)}\sqrt{u(u-1)(u-x)}} 
	\left[-x(x-1)u' +\frac{u(u-1)(t-x)}{t-u}\right]   
	\begin{pmatrix} 1 & 0 \cr 0 & -1 \cr\end{pmatrix}
\nonumber \\ & & {\hskip -4.0truemm}
	- \frac{(t-u)}{4\sqrt{t(t-1)(t-x)}}
	\left[
	\theta_\infty 
	\begin{pmatrix} 0 & 1 \cr  1 &  0 \cr\end{pmatrix}
	+ \sum_{j=1}^3 \theta_j 
	\begin{pmatrix} 0 & \sqrt{B_{j,+}} \cr \sqrt{B_{j,-}} & 0 \cr\end{pmatrix}
		\right].
\label{eqLM1}
\end{eqnarray}
The algebraic functions of $(u,x,t)$ appearing above are
\begin{eqnarray}
& & {\hskip -10.0truemm} 
%					
%F_{j,\pm}=\frac{1}{4 x(x-1)}
%\left[
% \frac{t(t-1)\D x-x(x-1)\D t}{\sqrt{t(t-1)(t-x)}}-4 A_{j,\pm} \D x
%\right] \sqrt{B_{j,\pm}} \sqrt{t-u},
%\nonumber \\ & & {\hskip -10.0truemm}
%
%write 0==(e21*u+e1-ej)
%  *(e21**3*sqrt(fprod(u,x))**2-(e21*(t+u)+2*e1+ej)*(e21*u+e1-ej)**2)
%  -ajval[_den];
%
%write 0==-(e21/4)*(sqrt(fprod(t,x))*(e21*u+e1-ej)**3
%               +e21**3*eps*sqrt(fprod(u,x))**3
%     -(e21*u+e1-ej)**2*eps*sqrt(fprod(u,x))*(e21*u+e1+2*ej))-ajval[_num];
%
A_{j,\pm}=\frac{\wp'(2 \omega T)-\wp'(\pm 2 \omega U+\omega_j)}{8 \sqrt{e_2-e_1}(\wp(2 \omega T)-\wp(\pm 2 \omega U+\omega_j)}
\nonumber \\ & & {\hskip -10.0truemm} \phantom{A_j}
=-\frac{1}{4 \alpha_j}
  \frac{\alpha_j^3 \sqrt{t(t-1)(t-x)}\mp \alpha_j^2\beta_j\sqrt{u(u-1)(u-x)}\pm (u(u-1)(u-x))^{3/2}}
       {u(u-1)(u-x)-\alpha_j^2\gamma_j},
\nonumber \\ & & {\hskip -10.0truemm}
%
%write 0==-(t-u)**2*(e21*u+e1-ej)-bjval[_den];
%write 0==let eps**2=1 in -e21*(sqrt(fprod(t,x))-eps*sqrt(fprod(u,x)))**2
%  +(t-u)**2*(e21*(t+u)+2*e1+ej)-bjval[_num];
%
B_{j,\pm}=\frac{\wp(2 \omega(\pm U+T))-e_j} {\wp(2 \omega U)-e_j}
=\frac{(\sqrt{t(t-1)(t-x)} \mp \sqrt{u(u-1)(u-x)})^2-\gamma_j(t-u)^2}{\alpha_j(t-u)^2},
\nonumber \\ & & {\hskip -10.0truemm}
\alpha_j=u  -\frac{  e_j- e_1}{e_2-e_1},\
 \beta_j=u  +\frac{2 e_j+ e_1}{e_2-e_1},\
\gamma_j=t+u+\frac{  e_j+2e_1}{e_2-e_1}\cdot
\end{eqnarray}

% ============================================================================
\subsubsection{Third order matrix Lax pair of Harnad}
% ============================================================================

Finally,
it is worth saying a few words on a nice third order matrix Lax pair 
\cite{Harnad1994},
associated to a three-degree of freedom Hamiltonian system.
This Lax pair, which has one Fuchsian singularity and one nonFuchsian,
admits a dual, second order matrix Lax pair 
defined in \cite[Eq.~(3.55), (3.61)]{Harnad1994},
which has the structure (\ref{eq-Matrix-Lax-pair})
and whose residues can be found in \cite[Eqs.~(65)--(68)]{CGM}.
Its identification to the present Lax pair
(\ref{eqLax-PVI-codim0-balanced-meromorphic})
could define a rational representation of the Hamilton variables
$(q_j,p_j)$, $j=1,2,3$,
instead of the algebraic ones obtained by identification
to the Lax pair of Jimbo and Miwa.
This could also help to solve the factorization problem 
of the three-wave system mentioned in \cite[\S 7.2]{CGM}.

% ============================================================================
\section{Quantum correspondence}
\label{sectionQuantum_correspondence}

In 1994 Suleimanov \cite{Suleimanov1994} 
noticed 
remarkable properties for $\PVI$,
which are inherited by all the lower Painlev\'e equations under the confluence.

\begin{enumerate}
\item
There exists a linear PDE of the parabolic type
\begin{eqnarray}
& & {\hskip -11.0truemm}
\left[x(x-1)(\partial_x+g_{\rm h}(x))-t(t-1)(t-x) \partial_t^2 
+v(x,t)  \right] \psi_{\rm h}=0,\
\label{eqheatxt} 
\end{eqnarray}
(the subscript h refers to the heat equation),
in which 
the potential $v$ is a rational function,
and
$g_{\rm h}$ an irrelevant arbitrary function,
both of them independent of the nonlinear field $u$.
This generalized heat equation,
which is defined up to the multiplication of
$\psi_{\rm h}$ by an arbitrary function of $(x,t,\theta_j)$,
can be normalized in different ways,
either affine in the four $\theta_j^2$'s \cite{Suleimanov1994},
or affine in ($\theta_0$, $\theta_1$, $\theta_x$, 
$\theta_\infty^2-(\theta_0+\theta_1+\theta_x-1)^2)$
\cite{N2009,CD2014}.
 
\item
There exists a representation of $\PVI$ by a classical Hamiltonian $\Hqp(q,p,x)$
with $q=u$,
different from the one of Malmquist \cite{MalmquistP6},
and there exists a quantization 
$q \to t$, $p \to \partial_t$, $\Hqp(q,p,x) \to \Hqp(t,\partial_t,x)$ 
(with an appropriate ordering of the noncommuting operators $t$, $\partial_t$)
allowing the identification of 
the generalized heat equation (\ref{eqheatxt})
to the time-dependent Schr\"odinger equation of quantum mechanics,
\begin{eqnarray}
& & {\hskip -11.0truemm}
\left[x(x-1)(\partial_x+g_{\rm h}(x))-\Hqp(t,\partial_t,x)\right] \psi_{\rm h}=0.
\label{eqheatxt-Quantum_correspondence} 
\end{eqnarray}

\item
This ``quantum correspondence'' extends \cite{ZZ2012PVI,ZZ2012PIPV} 
to the representation of $\PVI$ in elliptic coordinates $(X,T)$.
The heat equation (\ref{eqheatxt}) then takes the form \cite[Eq.~(5.21)]{ZZ2012PVI}
\begin{eqnarray}
& & {\hskip -11.0truemm}
\left[-a (\partial_X-G_{\rm H}(X))+\frac{1}{2}(\partial_T^2 +V(T,X,\lbrace \theta_j^2-1/4\rbrace))\right] \Psi_{\rm H}=0,\ % a?
\label{eqHeatXT} 
\end{eqnarray}
and the potential $V$ is deduced from the elliptic potential $V$ in (\ref{eqHamiltonien-wp})
by shifting each $\theta_j^2$ by $-1/4$.

\end{enumerate}

However, as pointed out in \cite[p.~4]{ZZ2012PIPV},
the quantum correspondence in rational coordinates 
(\ref{eqheatxt-Quantum_correspondence}) 
relies, at least for $\PVI$, $\PV$, $\PIV$ and $\PIII$ 
(see details in \cite{Suleimanov1994,Suleimanov2012}),
on a skilful, unexplained,
choice of the ordering of the products of $t$ and $\partial_t$.
Indeed, 
while the Hamiltonian (\ref{eqHamiltonien-wp}) in elliptic coordinates 
is the sum of a ``kinetic energy'' $P^2/2$ and a ``potential energy'' $V(Q,X)$,
this is no more the case in rational coordinates $(u,x,t)$. 
Moreover, 
if one uses a nonoptimal matrix Lax pair,
the derivation of the heat equation in elliptic coordinates
makes the computations rather involved \cite{ZZ2012PVI}.
\medskip

It is therefore necessary to give a direct, deterministic derivation of all these results.
Let us do that
only starting from the matrix Lax pair provided by the moving frame of Bonnet surfaces.

\textbf{First step}.
Since the heat equation is scalar,
one must first derive a scalar Lax pair from the matrix one.
Each of the four off-diagonal elements $M_{12}$, $M_{21}$ of 
(\ref{eqLax-PVI-codim0-unbalanced-holomorphic})
and
(\ref{eqLax-PVI-codim0-balanced-meromorphic})
possesses a single zero $t=f(u',u,x)$
(provided one sets $\auu=\pm \theta_\infty$ 
 in (\ref{eqLax-PVI-codim0-unbalanced-holomorphic})),
and each of these four zeroes obeys a $\PVI$ equation. 
These contiguous $\PVI$ are linked by birational transformations 
as sketched in \cite[Eq.~(4.4)]{LCM2003}.
The simplest of these elements is
\begin{eqnarray}
& & 
(\ref{eqLax-PVI-codim0-unbalanced-holomorphic}):\
M_{12}=\frac{t-u}{t(t-1)}\cdot
\end{eqnarray}
The elimination of anyone of the two components of the moving frame,
whether in  (\ref{eqLax-PVI-codim0-unbalanced-holomorphic})
or in
(\ref{eqLax-PVI-codim0-balanced-meromorphic}),
therefore generates the unique apparent singularity 
($t=u$ in the above example,
 $t=$ another $\PVI$ function in the three other cases)
which the scalar Lax pair must possess \cite[p.~219]{Poincare1883}.

If one denotes the two components of 
(\ref{eqLax-PVI-codim0-unbalanced-holomorphic}) and
(\ref{eqLax-PVI-codim0-balanced-meromorphic})
as, respectively,
$\psi_{j,\rm q}$ (q like quadratic in the $\theta_j$'s),
$\psi_{j,\rm m}$ (m like meromorphic in $\theta_\infty$),
then the scalar wave vector
\begin{eqnarray}
& & {\hskip -15.0 truemm}
\psi_{\rm d}=\sqrt{\frac{t(t-1)}{t-u}} \psi_{1,\rm q},\
\label{eqpsiq-to-psid}
\end{eqnarray} 
obeys the scalar Lax pair (\ref{eqFuchsScalarPDE})
of R.~Fuchs \cite{FuchsP6},
\begin{eqnarray}
& & {\hskip -10.0 truemm}
\left(\partial_t^2 + \frac{S}{2} \right) \psi_{\rm d}=0,\
%\label{eqGarnierScalarODE}
\left(\partial_x + C \partial_t  -\frac{C_t}{2} + g_{\rm d}(x,u,u')\right) \psi_{\rm d} = 0,\ 
\label{eqLaxd}
\end{eqnarray}
in which the arbitrary function $g_{\rm d}$,
which depends on $x,u(x),u'(x)$ but not on $t$,
will be later used to cancel various terms independent of $t$.

The coefficient $C$ is independent of the four $\theta_j$'s 
and the coefficient $S$ (the Schwarzian),
which has five double poles in $t$ (hence the notation $\psi_{\rm d}$)
has two nice properties which we will preserve throughout this section:
(i) it is an even function of the four $\theta_j$'s;
(ii) as noticed by Garnier \cite[p.~51]{GarnierThese},
it displays a remarkable symmetry between $u$ and $t$.
Indeed, if one introduces the potential function
\begin{eqnarray}
& & {\hskip -10.0 truemm}
V_{\rm G}(z,s)=\frac{1}{4}\left[-3 z
+(\theta_\infty^2-s) (z-x)
+(\theta_0     ^2-s)\left(\frac{x}{z}-1\right) \right.
\nonumber\\ & & {\hskip -10.0 truemm}\phantom{123456} \left.
+(\theta_1     ^2-s)\left(-\frac{x-1}{z-1}+1\right) 
+(\theta_x     ^2-s)\left(\frac{x(x-1)}{z-x}+2 x-1\right)\right],
\label{eqGarnier_Potential}
\end{eqnarray}
the dependence of $S$ on the $\theta_j$'s is only through
the difference $V_{\rm G}(u,s)-V_{\rm G}(t,s)$
in which the shift $s$ is unity,
\begin{eqnarray}
& & {\hskip -10.0 truemm}
C=-\displaystyle\frac{t(t-1)(u-x)}{x(x-1)(t-u)}\ccomma
  \quad
\label{eqC}
\\ & & {\hskip -10.0 truemm}
 \frac{S}{2}=
-\frac{3/4}{(t-u)^2}
- \frac{\beta_1 u' + \beta_0}{t(t-1)(t-u)}
-\frac{[(\beta_1 u')^2 - \beta_0^2] (u-x)}{u (u-1)t(t-1)(t-x)}
\nonumber\\ & & {\hskip -10.0 truemm}\phantom{123456}
+\frac{1}{t(t-1)(t-x)}\left[V_{\rm G}(u,1)-V_{\rm G}(t,1)\right],
%RC Corrigendum The P handbook p 210 (B.58)
\label{eqS}
\\ & & {\hskip -10.0 truemm}
\beta_1=\displaystyle-\frac{x (x-1)}{2 (u-x)}\ccomma
  \quad 
\beta_0=-u+\frac{1}{2}\cdot
%(2\alpha, -2\beta, 2\gamma, 1-2\delta) 
%=(\theta_\infty^2,\theta_0^2,\theta_1^2,\theta_x^2)
%=(4(a+b+c+d+1),4 a+1,4 b+1,4 c+1),
\end{eqnarray}

\textit{Remark}.
In the correspondence matrix-scalar,
defined by (\ref{eqpsiq-to-psim}) and
\begin{eqnarray}
& & {\hskip -15.0 truemm}
\psi_{1,\rm q}=\sqrt{\frac{t-u}{t(t-1)}}\psi_{\rm d},\
\psi_{2,\rm q}=
\left(
 \frac{x(x-1)}{u-x} \left(\partial_x 
 -\Theta_x \frac{1}{2(t-x)}\right)+\frac{\auu}{2}
\right)\psi_{1,\rm q},
\end{eqnarray} 
only $\psi_{1,\rm q}$ has a simple dependence on $\psi_{\rm d}$,
and the correspondence between the second component $\psi_{2,\rm q}$ and $\psi_{\rm d}$ 
is indeed quite difficult \cite{LCM2003,C2006Kyoto}
to find simply by some good guess.

% ============================================================================

\textbf{Second step}.
This is the
elimination of $u$ between the two equations of the scalar Lax pair (\ref{eqLaxd}).
It is realized \cite{CD2014} by 
the linear combination
$x(x-1)(\partial_x+\dots)-t(t-1)(t-x) (\partial_t^2+\dots)$ of the two scalar
equations (\ref{eqLaxd}),
followed by
a change of the wave function $\psi_{\rm d}$ 
and a suitable choice of the arbitrary function $g_{\rm d}(x,u,u')$.
This change is essentially 
$\psi_{\rm d}= (t-u)^{-1/2}\psi_{\rm h}$
but, because of the freedom $\psi_{\rm h} \to f(x,t)\psi_{\rm h}$
with $f$ independent of $u$,
it is more convenient to define it as
\begin{eqnarray}
& & {\hskip -11.0truemm}
\psi_{\rm d}= (t-u)^{-1/2} t^{k_0/2} (t-1)^{k_1/2}(t-x)^{k_x+1/2} \psi_{\rm h},
\label{eqpsid-to-psih}
\end{eqnarray}
with $k_0$, $k_1$, $k_x$ adjustable constants independent of the $\theta_j$'s.
The transformed Lax pair thus retains the parity in the $\theta_j$'s.

The crucial question at this point is to find the Hamiltonian $\Hqp(q,p,x)$
whose quantization $\Hqp(t,\partial_t,x)$ is unambiguous and
succeeds to describe the resulting scalar heat equation 
(\ref{eqheatxt-Quantum_correspondence}) 
for the $\psi_{\rm h}$ defined in (\ref{eqpsid-to-psih}).
Indeed, the Hamiltonian description of $\PVI$ 
(at least in rational coordinates) is not unique.
The three Hamiltonians we are aware of
(Malmquist  \cite{MalmquistP6},
 Suleimanov \cite{Suleimanov1994},
 Tsegel'nik \cite{T2007-H-PVI})
have different properties:
polynomial in $q$ (Malmquist),
even functions of $p$ (Malmquist, Suleimanov),
even functions of the four $\theta_j$'s (Suleimanov, Tsegel'nik),
but none of these properties is relevant,
and this is a fourth property which dictates the Hamiltonian.

Indeed, in order to prove the Painlev\'e property of $\PVI$,
Painlev\'e built \cite[p.~26]{PaiActa} \cite[Eq.~(3)]{PaiCRAS1906} 
four rational functions of $u'$, $u$, $x$
having as only movable singularities 
two movable simple poles of residue unity 
(reached when $u \sim \pm x(x-1) \theta_\infty^{-1} (x-x_{0,\pm})^{-1}$
in the expression below,
and similarly for the three other rational functions),
\begin{eqnarray}
%function tauPVIP(u,x) {declare u function;return(
%   (1/2)*x*xm1(x)*deriv(u(x),x)**2/(u(x)*um1(x)*umx(x))
%    -deriv(u(x),x)/umx(x)
%  +(1/2)/x/xm1(x)*(
%   ti**2*(-u(x)+1/2)+t0**2*(-x/u(x)+1/2)+t1**2*(xm1(x)/um1(x)-1/2)
% +(tx**2-1)*(-x*xm1(x)/umx(x)+1/2-x)));};
& & {\hskip -2.0 truemm}
\Dlogtau_{\rm VI,P}=\frac{x(x-1) {u'}^2}{2 u(u-1)(u-x)} - \frac{u'}{u-x}
\label{eqtauPVIP}
\\ & & 
+\frac{1}{2 x(x-1)}
\left[
 \theta_\infty^2 \left(\frac{1}{2}-u\right)        
+\theta_0^2      \left(\frac{1}{2}-\frac{x}{u}\right)
\right. \nonumber\\ & & \phantom{1234567890}\left.
+ \theta_1^2     \left(\frac{x-1}{u-1} -\frac{1}{2}\right)
+(\theta_x^2-1)  \left(\frac{1}{2}-x-\frac{x(x-1)}{u-x}\right)
-x+1
\right],
\nonumber
\end{eqnarray}
and this is the Hamiltonian associated to this tau-function
which correctly defines the quantum correspondence.
The Hamiltonians of Malmquist and Suleimanov
are associated to a different tau-function,
which is the one 
built by Chazy \cite[expression $t$ page 341]{ChazyThese}
and whose logarithmic derivative has only one (instead of two)
movable simple pole of residue unity
(reached when $u \sim x(x-1) \theta_\infty^{-1} (x-x_0)^{-1}$).

The Hamiltonian \cite{T2007-H-PVI} associated to (\ref{eqtauPVIP}),
\begin{eqnarray}
& & {\hskip -15.0 truemm}
\left\lbrace 
\begin{array}{ll}
\displaystyle{
%function HamVIT(q,p,x) (1/aT)*1/x/xm1(x)*
% (q*(q-1)*(q-x)*aT**2*p**2+q*(q-1)*aT*p
% +(1/4)*(0
% +(ti**2-1)*(-q+1/2)+t0**2*(-x/q+1/2)+t1**2*(xm1(x)/(q-1)-1/2)
%  +tx**2*(-x*xm1(x)/(q-x)+1/2-x)
%  +2*x-1));
\Hqp_{\rm VI,T}(q,p,x)=\frac{1}{\aT x(x-1)}
\left\lbrack
 q(q-1)(q-x) \aT^2 p^2 + q(q-1) \aT p
\right.
}\\ \displaystyle{
\phantom{12345678}\left.
+\frac{1}{4}\left(
 (\theta_\infty^2-1) \left(\frac{1}{2}-q\right) 
+ \theta_0^2         \left(\frac{1}{2}-\frac{x}{q}\right)
\right. \right.}\\ \displaystyle{\phantom{1234567890}\left.\left.
+ \theta_1^2         \left(-\frac{1}{2}+\frac{x-1}{u-1} \right)
+ \theta_x^2         \left(\frac{1}{2}-x-\frac{x(x-1)}{u-x}\right)
+2 x-1
            \right)\right\rbrack,
}\\ \displaystyle{
q=u,\
%pvalT=aT**(-1)*(x*xm1(x)*'(u,1)(x)/u(x)/um1(x)/umx(x)-1/umx(x))/2;
p=\frac{1}{2 \aT}\left(\frac{x(x-1) u'}{u(u-1)(u-x)}-\frac{1}{q-x}\right),
}
\end{array}
\right.
\label{eqHamVIT}  
\end{eqnarray} 
($\aT$ being a constant of normalization),
only differs from (\ref{eqtauPVIP})
by an additive term which reflects the singling out of one singular point among four,
\begin{eqnarray}
& & {\hskip -15.0 truemm}
% tauPVIP(u,x)-2*aT*HamVIT(q,p,x)+deriv(log(u(x)-x),x);
\Dlogtau_{\rm VI,P} - 2 \aT \Hqp_{\rm VI,T}(q,p,x) + \frac{\D }{\D x} \log (q-x)=0.
\end{eqnarray}

Then, if one defines the quantum Hamiltonian as
\begin{eqnarray}
& & {\hskip -15.0 truemm}
\forall \psi:\ 
\Hqp_{\rm VI,T}(t,\partial_t,x) \psi
=\aT \frac{t(t-1)(t-x)}{\aT^2 x(x-1)} \partial_t^2 \psi
 +   \frac{t(t-1)     }{\aT   x(x-1)} \partial_t   \psi
 + (\partial_t^0 \hbox{ terms}) \psi,
\label{eqQuantumHamVI}
\end{eqnarray}
% k0=0,k1=0,kx=kp,si=-1+2*kp,s0=-1,s1=-1,sx=-1,c0=0,kp**2=1 inactive;
and chooses the above adjustable parameters as $k_0=0, k_1=0, k_x=1$,
the scalar Lax pair for the $\psi_{\rm h}$ defined in (\ref{eqpsid-to-psih}) 
can be written as
\begin{eqnarray}
& & {\hskip -15.0 truemm}
\left\lbrace 
\begin{array}{ll}
\displaystyle{
% AMP Laxh2
\left[\partial_x-\aT\Hqp_{\rm VI,T}(t,\partial_t,x,\theta_{j}^2+s_j)+g_{\rm h}(x)
\right] \psi_{\rm h}=0,\
}\\ \displaystyle{
% AMP Laxh1
\left\lbrack\partial_x 
-\displaystyle\frac{t(t-1)(u-x)}{x(x-1)(t-u)} \partial_t 
-\frac{1}{2}\Dlogtau_{\rm VI,P}
+ g_{\rm h}(x)
-\frac{3}{4 x}-\frac{3}{4 (x-1)}
\right.}\\ \displaystyle{\left.\phantom{12}
+\frac{x(x-1)u'+u(u-1)(u-x)\left(\displaystyle\frac{1}{u}+ \frac{1}{u-1}- \frac{2}{u-x} \right)}
      {2 x(x-1)(t-u)}
\right\rbrack\psi_{\rm h} = 0,\
}\\ \displaystyle{
%let '(G,1)(x)=-frab(x)+(gh(x)-gd(x))/x/xm1(x)+shifthd in answer;
%-shifthd-(1/2)*tauPVIP(u,x)+frab(x)+(-kp-2)/4/x+(-kp-2)/4/xm1(x);
g_{\rm d}-g_{\rm h}=
 -\frac{1}{2}\Dlogtau_{\rm VI,P}-\frac{3}{4 x}-\frac{3}{4 (x-1)},
}\\ \displaystyle{
s_\infty=1,\ s_0=s_1=s_x=-1,
}
\end{array}
\right.
\label{eqLaxhuxt}
\end{eqnarray} 
and the quantum correspondence is realized by taking $g_{\rm h}$ as the arbitrary function.
The reason why only three $\theta_{j}^2$ display the same shift
is a consequence of 
the necessity to single out one of the four singular points
in order to define a tau-function.

% ============================================================================

\textbf{Third step}.
The conversion of this Lax pair to elliptic coordinates $(U,X,T)$
is performed in a systematic way following the guidelines
and the formulae of Halphen recalled 
in Appendix \ref{AppendixElliptic3var}.
In order that the heat equation in elliptic coordinates
takes the normalized form (\ref{eqHeatXT}) (i.e.~without $\partial_T$ term),
the wave function $\Psi_{\rm H}$ must be
\begin{eqnarray}
& & {\hskip -15.0 truemm}
\left\lbrace 
\begin{array}{ll}
\displaystyle{
\Psi_{\rm H}=(t(t-1))^{-1/4} (t-x)^{3/4} e^F \psi_{\rm h},\
}\\ \displaystyle{
\D F=
\frac{W(t,x)}{2(t(t-1)(t-x))^{1/2}} \D t 
+\frac{G_{\rm H}(X)}{x(x-1)} \D x
}\\ \displaystyle{
\phantom{12}
+\frac{-2 (t(t-1)(t-x))^{1/2} W(t,x)-(t-x) W^2(t,x)-(t-x)^2}{4 x(x-1)(t-x)} \D x,
}
\end{array}
\right.
\label{eqLaxUXTPsi}
\end{eqnarray} 
in which $G_{\rm H}(X)$ is an arbitrary function.
As expected, the link between the wave function 
$\Psi_{\rm H}$ of the heat equation in elliptic coordinates
and the wave function $\psi_{1,\rm q}$ of the first canonical form
of the matrix Lax pair of $\PVI$
does not involve the apparent singularity $t=u$,
\begin{eqnarray}
& & {\hskip -15.0 truemm}
\Psi_{\rm H}=(t(t-1))^{1/4} (t-x)^{-3/4} e^F \psi_{1,\rm q}.
\end{eqnarray} 

In the elliptic coordinates, 
the Lax pair (\ref{eqLaxhuxt}) then becomes
\begin{eqnarray}
& & {\hskip -15.0 truemm}
\left\lbrace 
\begin{array}{ll}
\displaystyle{
% '(PsiH,1,0)(X,T)*part10
%+'(PsiH,0,1)(X,T)*(part01/'(wpg,1,0,0)(zT,wg2,wg3)+wzetag(zT,wg2,wg3)-2*eta*T)
%    *omega
%+  PsiH     (X,T)*(0
% +part002*(du/dx)**2*omega**2/e21
% +part001*(du/dx)   *omega**2/e21
% +part0002*W(t,x)**2*omega**2/e21/12
% +part0001*W(t,x)*sqrtpt   *omega**2/e21/12
% +part00001*V3(x)   *omega**2/e21/12
% +part00000         *omega**2/e21/12
\left[2 a_X \pi^2 \partial_X-\frac{1}{2}G_{\rm H}(X)+\frac{1}{2}\partial_T^2 
 -\frac{1}{2}(2\omega)^2 
 \sum_{j=\infty,0,1,x} \left(\theta_j^2-\frac{1}{4} \right)\wp(2\omega T+\omega_j)\right] \Psi_{\rm H}=0,\
}\\ \displaystyle{
\left\lbrack 
 a_X \pi^2 \partial_X 
+\omega\left\lbrack 
\frac{1}{2}\frac{\wp'(2 \omega T)}{\wp(2\omega T)-\wp(2\omega U)}+\zeta(2\omega T)-2 \eta T
\right\rbrack\partial_T 
+ (e_2-e_1)\omega^2 G_{\rm H}(X)
\right.}\\ \displaystyle{\left.\phantom{12}
\right.}\\ \displaystyle{\left.\phantom{12}
-(e_2-e_1)\omega^2[
  \theta_\infty^2    \wp(2\omega U)
 +\theta_0     ^2    \wp(2\omega U+\omega_1)
 +\theta_1     ^2    \wp(2\omega U+\omega_2)
+(\theta_x     ^2-1) \wp(2\omega U+\omega_3)
]
\right.}\\ \displaystyle{\left.\phantom{12}
\right.}\\ \displaystyle{\left.\phantom{12}
+\frac{(2\omega)^2 (e_3-e_2)^2(e_1-e_3)^2}{{\wp'}^2(2 \omega U)}
 \left(\frac{\D u}{\D x}\right)^2
+\frac{2 \omega^2(e_3-e_2)(e_1-e_3)(\wp(2\omega T)-e_3)}
   {(\wp(2\omega T)-\wp(2\omega U))(\wp(2\omega U)-e_3)}
 \frac{\D u}{\D x}
\right.}\\ \displaystyle{\left.\phantom{12}
\right.}\\ \displaystyle{\left.\phantom{12}
- \omega^2 (\zeta(2\omega T)-2 \eta T)^2 % W^2(t,x) term
-\omega^2\frac{(\zeta(2\omega T)-2 \eta T)\wp'(2\omega T)}{\wp(2\omega T)-\wp(2\omega U)} 
-\omega (\eta+e_3 \omega) % V_3(x) term
\right.}\\ \displaystyle{\left.\phantom{12}
\right.}\\ \displaystyle{\left.\phantom{12}
-\omega^2 \frac{(\wp(2\omega T)-e_3)^2-(e_3-e_1)(e_3-e_2)}{\wp(2\omega T)-\wp(2\omega U)}
\right\rbrack \Psi_{\rm H}=0,\
}
\end{array}
\right.
\label{eqLaxUXT}
\end{eqnarray}
in which $\D u/ \D x$ should be replaced by the expression (\ref{eqdusurdx}).

The reduction $\partial_X=0$ of the remarkable parabolic PDE (\ref{eqLaxUXT})${}_1$
is a generalization to nonconstant half-periods
of the equation introduced by Darboux \cite{Darboux1882-Four-wp},
\begin{eqnarray}
& & {\hskip -15.0 truemm}
\left[\frac{\D^2}{\D T^2} - \lambda
 -(2\omega)^2 \sum_{j=\infty,0,1,x} n_j(n_j+1)\wp(2\omega T+\omega_j)\right] \Psi=0,\
\label{eqDarboux}
\end{eqnarray}
integrated for arbitrary complex $\lambda$ and integer $n_j$'s by de Sparre \cite{deSparre1883},
and rediscovered by Treibich and Verdier more than one century later.
Its writing in rational coordinates is 
identical to the Heun equation \cite{Heun}. % Give the formula, see 2016.Chiang_Tsang_Chiang slide 25
For a modern account on this Darboux ODE, see \cite{RonveauxBook,VeselovDarboux}.

\textit{Remark}.
The shifts $-1/4$ in (\ref{eqLaxUXT})${}_1$ are a direct consequence of the identity
$n_j(n_j+1)=(n_j+1/2)^2-1/4$.

The confluence to the lower Painlev\'e equations of the results presented in 
this section can be found in Appendix \ref{AppendixConfluence}.

% ============================================================================
\section{Generalized Bonnet surfaces}
\label{sectionBonnet_plus}
% ============================================================================

Given the previous results,
it is natural to ask whether there exist analytic surfaces represented by the full $\PVI$,
which would therefore generalize Bonnet surfaces.
The answer is indeed positive.

The matrix Lax pair
(\ref{eqLax-PVI-codim0-unbalanced-holomorphic})
can be converted back to the moving frame (\ref{eqR3R4_Moving_frame_order2})
of some surface by solving the six scalar equations
\begin{eqnarray}
& & 
\matU P ^{-1}\D z + \matV P ^{-1}\D \barz=L \D x + M \D t,\
P=
\begin{pmatrix} G & 0 \cr 0 & 1/G \cr\end{pmatrix},
\label{eqBack-to-HyperBonnet}
\end{eqnarray} 
for the five unknowns $e^\metric$, $H$, $Q$, $\barQ$, $G^2$
(indeed, $c$ scales out because of (\ref{eqScaling})),
thus defining an extrapolation of the Bonnet moving frame (\ref{eqdpsiBonshort}) 
to arbitrary values of the four $\theta_j$'s.
%When $c$ is zero, one only finds the solution of Bonnet.
%
%When $c$ is nonzero,
If one defines $f_{jk}$ and $g_{jk}$ by
\begin{eqnarray}
& & {\hskip -16.0truemm}
L_{jk} \D x + M_{jk} \D t = f_{jk}(x,t) \D z + g_{jk}(x,t) \D \barz,
\end{eqnarray} 
this system of six scalar equations, % defined by (\ref{eqBack-to-HyperBonnet})
\begin{eqnarray}
& & {\hskip -15.0 truemm}
\left\lbrace 
\begin{array}{ll}
\displaystyle{
    Q e^{-\metric/2} G^{-2}= -4 \cz (t-x),\
\barQ e^{-\metric/2} G^2   = g_{21},
}\\ \displaystyle{
(H+c) e^{ \metric/2} G^2   = 2 f_{21},\
(H-c) e^{ \metric/2} G^{-2}= 8 \cz (u-x),
}\\ \displaystyle{
\D \log (e^{ \metric} G^4)= -f_{11} \D z
 +2 \cz \left[ \auu (u-x) - \Theta_x \frac{x(x-1)}{t-x}\right] \D \barz, % \auu=?
}
\end{array}
\right.
\label{eqBack-to-GC}
\end{eqnarray} 
is equivalent to
\begin{eqnarray}
& & {\hskip -16.0truemm}
\left\lbrace 
\begin{array}{ll}
\displaystyle{
G^4=-4 \frac{f_{21}}{4 \cz (u-x)} \frac{H-c}{H+c}\ccomma\
e^{\metric}=- \frac{16 \cz (u-x) f_{21}}{H^2-c^2}\ccomma\
}\\ \displaystyle{
    Q=- \frac{8 \cz (t-x) f_{21}}{H+c}\ccomma\		
\barQ=- \frac{8 \cz (u-x) g_{21}}{H-c}\ccomma\
}\\ \displaystyle{
\D H=(H-c) A \D z +(H+c) B\D \barz,
}\\ \displaystyle{
A(x,t)=-2 f_{11} + 4 x(x-1) \frac{u'-1}{u-x}\ccomma\
}\\ \displaystyle{
B(x,t)=4 \cz \left[\Theta_x \frac{x(x-1)}{t-x} - \auu (u-x) + x(x-1) \partial_x \log(f_{21}) \right],
}
\end{array}
\right.
\label{eq-HyperBonnet-guQdH}
\end{eqnarray} 
and the condition $\D^2 H=0$,
\begin{eqnarray}
& & 
\ \ (H-c)\left[A B - 4 \cz x(x-1) A_x\right]
\nonumber\\ & & 
-(H+c)\left[A B - 4 \cz t(t-1) B_t - 4 \cz x(x-1) B_x\right]=0,
\label{eq-HyperBonnet-H}
\end{eqnarray} 
admits three solutions
\begin{eqnarray}
& & {\hskip -16.0truemm}
\left\lbrace 
\begin{array}{ll}
\displaystyle{
({\rm A})\ c\not=0,\ (\theta_\infty,\Theta_x)\not=(-1,-1):\ 
\frac{H}{c}=\hbox{ the unique solution of } (\ref{eq-HyperBonnet-H}),
}\\ \displaystyle{
({\rm B})\ c=0,\ \theta_\infty=\Theta_x\not=-1,\ \theta_1^2=\theta_0^2:\ 
H=\hbox{ any integral of } (\ref{eq-HyperBonnet-guQdH})_3,
}\\ \displaystyle{
({\rm C})\ c \hbox{ arbitrary},\ (\theta_\infty,\Theta_x)=(-1,-1):\ 
H=\hbox{ any integral of } (\ref{eq-HyperBonnet-guQdH})_3.
}
\end{array}
\right.
\label{eq-HyperBonnet-3sol}
\end{eqnarray}

The generic solution $({\rm A})$
\begin{eqnarray}
& & {\hskip -16.0truemm}
c\not=0,\ (\theta_\infty,\Theta_x)\not=(-1,-1):\ 
\frac{H}{c}
=\frac{(\theta_\infty-\Theta_x) P_5+(\theta_\infty+\Theta_x-2) P_3}
      {(\theta_\infty-\Theta_x) P_4+(\theta_\infty+\Theta_x-2) P_5}\ccomma
\nonumber\\ & &	{\hskip -16.0truemm}
\phantom{c\not=0,\ (\theta_\infty,\Theta_x)\not=(-1,-1)12}
\frac{H-c}{H+c}
=\frac{P_5} {x(x-1) P_3 \Dlogtau_{\rm VI,M}}\ccomma			
\label{eq-HyperBonnet-sol}
\end{eqnarray}
where the $P_n$'s, all different, are polynomials of 
$u'$, $u$, $x$, $t$, $\theta_\infty$, $\theta_0^2$, $\theta_1^2$, $\Theta_x$ of degree $n$ in $u'$,
defines an extrapolation of Bonnet surfaces
to more general surfaces in $\Rcubec$ represented by the generic $\PVI$,
in which,
as a consequence of (\ref{eqScaling}),
the nonzero value of $c$ is arbitrary and independent of the $\theta_j$'s.
Its Bonnet limit is, by construction,
\begin{eqnarray}
& & {\hskip -16.0truemm}
\lim_{\theta_\infty \to 0,\Theta_x \to 0}
\frac{H}{c}
=\frac{8 \cz x(x-1)\Dlogtau_{\rm VI,M}}{\cz(\theta_1^2-\theta_0^2)}\ccomma
\end{eqnarray}
but for arbitrary $\theta_j$'s the value of $H$ is different from
$8 \cz x(x-1)\Dlogtau_{\rm VI,M}$ and does depend on $t$.

As to the two nongeneric solutions $({\rm B})$ and $({\rm C})$,
in which $\D^2 H$ is identically zero,
they are not essentially different from the solution of Bonnet.
For instance, in the third solution $({\rm C})$,
two $\theta_j$'s are the same as those of Bonnet
and the two others are shifted by $\pm 1$,
therefore this third solution is the Schlesinger transform
\cite{SchlesingerP6,Garnier1951} %\cite[Eq.~(28)]{Garnier1951}
of the Bonnet solution,
and we leave it to the interested reader to establish the explicit expressions
for $H$.

\textit{Remark}.
One can similarly define an extrapolation of harmonic inverse mean curvature surfaces
to the full $\PVI$.
It is sufficient to first transpose the moving frame and therefore,
instead of solving (\ref{eqBack-to-HyperBonnet}), to solve
\begin{eqnarray}
& & 
{}^{t}\matU P ^{-1}\D z + {}^{t}\matV P ^{-1}\D \barz=L \D x + M \D t,\
P=
\begin{pmatrix} G & 0 \cr 0 & 1/G \cr
\end{pmatrix}.
\label{eqBack-to-HyperHIMC}
\end{eqnarray} 

% ==========================================================================
\section{Conclusion}

The present results are threefold:
(i) the natural Lax pair of $\PVI$;
(ii) a rigorous derivation of the quantum correspondence of $\PVI$;
(iii) an extension of Bonnet surfaces to two more parameters,
thus matching the completeness of $\PVI$ and the completeness of Gauss-Codazzi equations.

In future work, we plan two directions of research.

(i) To find a geometric characterization of these extended Bonnet surfaces.

(ii) To similarly improve the discrete matrix Lax pair
of $\qPVI$ introduced by Jimbo and Sakai \cite{JS1996},
i.e.~to remove its meromorphic dependence on a fixed parameter of this $\qPVI$.

% =======================================================================
\begin{acknowledgments}

This work was partially funded by the Hong Kong
GRF grant HKU 703313P and GRF grant 17301115. 
We gladly thank
l'Unit\'e mixte internationale UMI 3457 du Centre de recherches math\'ematiques de 
l'Universit\'e de Montr\'eal
for financial support.

\end{acknowledgments}

% =======================================================================
\appendix

% ============================================================================
\section{Conversion between rational and elliptic coordinates}
% ============================================================================
\label{AppendixElliptic3var}

% This appendix contains only $\Omega$, never $X$. Search for ``X'' to check

This Appendix has two guidelines.
The first one is to
define the Weierstrass functions $\wp$, $\zeta$, $\sigma$ 
as functions of three arguments $(z,g_2,g_3)$,
not two arguments $(z \vert \tau)$ by setting one period to unity.
The reason is that $\wp(z,g_2,g_3)$ is a homogeneous function,
thus making all formulae homogeneous and therefore easy to check.
Our second guideline is to never deal with partial derivatives,
always with differentials, in order to avoid thinking about which depends on what.
The reference is, naturally, the first volume of Halphen 
\cite[t.~I Chap.~IX--X]{HalphenTraite}.

The ratio of the two periods $2 \omega$, $2 \omega'$
and the discriminant are respectively denoted
$\Omega$ and $\Delta$ \cite[t.~I p.~321]{HalphenTraite},
\begin{eqnarray}
& &
% Halphen I p. 321 l. 3
\Omega=i \pi \frac{\omega'}{\omega},\
\Delta=g_2^3-27 g_3^2.
\label{eqOmegaDelta}
\end{eqnarray}

The transformation (\ref{eqdefUXT})--(\ref{eqdefuxt}) between rational and elliptic coordinates
is equivalently defined as
\begin{eqnarray}
& & {\hskip -15.0 truemm}
(u     ,0,1,x)=(\wp(2 \omega U),e_1,e_2,e_3),\
\Omega=i \pi \frac{\omega'}{\omega}\cdot
\end{eqnarray} 

In order to obtain (\ref{eqPVIUX}), it is sufficient to establish the differentials of $u$, $x$,
$\D u / \D x$ in terms of $\D U$, $\D \Omega$, $\D(\D U / \D \Omega)$,
then to substitute these values in (\ref{eqPVI}).

Since  $x$, $u$ and $\Omega,U$ have no dimension,
it is convenient to replace the triplet $(2 \omega U,g_2,g_3)$ of the arguments of $\wp$
by a triplet containing two dimensionless variables,
for instance $(U,\Omega,\Delta)$,
whose Jacobian
\begin{eqnarray}
& & 
\frac{D(2 \omega U,g_2,g_3 )}{D(U,\Omega,\Delta)}=2 \omega \frac{2 \omega^2}{9 \pi^2}
\end{eqnarray}
never vanishes.
The differential of a dimensionless variable will therefore have no contribution of
$\D \Delta$,
without the need for assuming the period $2 \omega$ to be unity.

If one denotes $(z,g_2,g_3)$ the three arguments of 
$\wp$, $\zeta$, $\sigma$,
and $L$ the dimension of the first argument $z$,
the various variables have the dimensions
\begin{eqnarray}
& & {\hskip -13.0 truemm}
[z]=[\omega]=[\sigma]=L,\
[\wp]=[e_\alpha]=L^{-2},\
[\zeta]=[\eta]=L^{-1},\
\nonumber\\ & &
[g_2]=L^{-4},\
[g_3]=L^{-6},\
[\Delta]=L^{-12}.
\end{eqnarray}

Abbreviating $\wp(z,g_2,g_3)$ (resp.~$\zeta(z,g_2,g_3)$, $\sigma(z,g_2,g_3)$) 
as $\wp$ (resp.~$\zeta$, $\sigma$),
and denoting by a quote $'$ the derivative with respect to the first argument,
the only necessary formulae, apart (\ref{eqOmegaDelta}), are the following
\cite[t.~I Chap.~IX--X]{HalphenTraite}\footnote{
Erratum. In formula 18.6.23 of Abramowitz and Stegun \cite{AbramowitzStegun}, 
the last $g_2$ should be $g_3$. 
},
\begin{eqnarray}
& & {\hskip -9.0truemm}
{\wp'}^2=4 \wp^3-g_2 \wp - g_3=4(\wp-e_1)(\wp-e_2)(\wp-e_3),\
\zeta'=-\wp,\ \sigma'=\sigma \zeta,\ 
\label{eqwp}
\\ & & {\hskip -9.0truemm}
\D \sigma
=\sigma \zeta \D z
+ \frac{1}{\Delta}
  \left( 
     -\frac{9}{4} g_3 \sigma'' +\frac{1}{4} g_2^2 z \sigma'
     -\left( \frac{1}{4} g_2^2 +\frac{3}{16} g_2 g_3 z^2 \right) \sigma
  \right) \D g_2
\nonumber\\ & & {\hskip -9.0truemm} \phantom{\D \sigma=\sigma \zeta \D z}
+ \frac{1}{\Delta}
 \left( 
   \frac{3}{2} g_2 \sigma'' -\frac{9}{2} g_3 z \sigma'
    + \left( \frac{9}{2} g_3 +\frac{1}{8} g_2^2 z^2 \right) \sigma
 \right) \D g_3,
\\ & & {\hskip -9.0truemm}
% -------------------------------------------------------------
\zeta(\omega,g_2,g_3)=\eta,\ 
\zeta(\omega',g_2,g_3)=\eta',\ 
% -------------------------------------------------------------
%  H vol I p 309 eq (40) = p 260 
\eta \omega'-\eta' \omega=i \frac{\pi}{2},\ 
\label{eqomega-eta}
\\ & & {\hskip -9.0truemm}
% -------------------------------------------------------------
% H vol I p 307 eq (37) 
\D \omega=
  \frac{1}{\Delta}\left(-\frac{1}{4} g_2^2 \omega +\frac{9}{2} g_3 \eta \right) \D g_2
+ \frac{1}{\Delta}\left( \frac{9}{2} g_3   \omega -          3 g_2 \eta \right) \D g_3,
\label{eqdomega}
%eqdomegap=let rprime in eqdomega;
\\ & & {\hskip -9.0truemm}
% -------------------------------------------------------------
% H vol I p 308 eq (39) 
\D \eta=
  \frac{1}{\Delta}\left(-\frac{3}{8} g_2 g_3 \omega +\frac{1}{4} g_2^2 \eta \right)  \D g_2
 +\frac{1}{\Delta}\left(\frac{1}{4} g_2^2    \omega -\frac{9}{2} g_3   \eta \right) \D g_3.
	\label{eqdeta}
%eqdetap=let rprime in eqdeta;
\end{eqnarray}
In particular, 
the expressions of $\D \zeta$ and $\D \wp$
result from $\D \sigma$ by action of the operator $'$
(we hope that no confusion occurs with $\omega'$),
which commutes with the operator $\D$.
To this list one should add the transformed of
(\ref{eqdomega})--(\ref{eqdeta})
under $(\omega,\eta) \to (\omega',\eta')$.

{}From the above formulae, one deduces the quite useful formula,
\begin{eqnarray}
& &
\D e_\alpha=\frac{e_\alpha \D g_2 + \D g_3}{12 e_\alpha^2-g_2}\ccomma
%\D \Delta=3 g_2^2 \D g_2 -54 g_3 \D g_3,\
%\D \Omega=-\pi^2 \left(-\frac{9}{4} g_3 \D g_2 +\frac{3}{2} g_2 \D g_3 \right) \Delta^{-1},
\end{eqnarray}
together with the formulae of the change of variables
\begin{eqnarray}
& & {\hskip -9.0truemm}
\D g_2=\frac{1}{\Delta}\left(           -12 g_3 \frac{\D \Omega}{\pi^2} + \frac{g_2}{3}\D \Delta\right),\
\D g_3=\frac{1}{\Delta}\left(-\frac{2}{3} g_2^2 \frac{\D \Omega}{\pi^2} + \frac{g_3}{2}\D \Delta\right),\
\\ & & {\hskip -9.0truemm}
\D \omega=2 \eta\omega^2 \frac{\D \Omega}{ \pi^2} - \frac{\omega}{12 \Delta} \D \Delta,\
\D \eta  = -g_2 \omega^3 \frac{\D \Omega}{6\pi^2} + \frac{\eta}  {12 \Delta} \D \Delta.
\end{eqnarray}

% ============================================================================
\subsection{From rational to elliptic coordinates}
% ============================================================================
\label{AppendixElliptic3var-rational-to-elliptic}

Given the definitions (\ref{eqdefuxt}) and now taking $z=2 \omega U$,
the differentials $\D x$ and $\D u$ are then linear forms of
$\D U$, $\D \Omega$ independent of $\D \Delta$,
\begin{eqnarray}
& &
\left\lbrace 
\begin{array}{ll}
\displaystyle{
\D x=-\frac{4 \omega^2 (e_3-e_1)(e_3-e_2)}{\pi^2 (e_2-e_1)} \D \Omega,
}\\ \displaystyle{
\D u=\frac{2 \omega \wp' }{(e_2-e_1)} \D U
+ \left[(\zeta - 2 \eta U)\wp'+2(\wp-e_1)(\wp-e_2)\right]
\frac{2 \omega^2 }{\pi^2 (e_2-e_1)} \D \Omega,
}
\end{array}
\right. 
\label{eqdxdu}
\end{eqnarray}
hence the value of $\D u / \D x$, 
\begin{eqnarray}
& & {\hskip -18.0truemm}
\frac{\D u}{\D x}=\frac{1}{2 \omega (e_3-e_1)(e_3-e_2)}
\left\lbrack
\pi^2 \wp'\frac{\D U}{\D \Omega} 
+ \left((\zeta - 2 \eta U)\wp'+2(\wp-e_1)(\wp-e_2)\right)\omega
\right\rbrack.
\label{eqdusurdx}
\end{eqnarray}

The differential of $\D u / \D x$ is similarly expressed as a linear form
of $\D U$, $\D \Omega$, $\D (\D U / \D \Omega)$, 
and one obtains
\begin{eqnarray}
& & {\hskip -7.0truemm}
\frac{\D^2 u}{\D x^2}=\frac{e_2-e_1}{8 \omega^3 (e_3-e_1)^2(e_3-e_2)^2}
\left\lbrack
\pi^4 \wp'\frac{\D^2 U}{\D \Omega^2} 
+ 2 \pi^4 \omega \wp'' \left(\frac{\D U}{\D \Omega} \right)^2
\right.\nonumber \\ & & {\hskip -9.0truemm} \phantom{1234567890123456789}\left.
- \pi^2 \omega^2 \left\lbrace 12 \wp \wp'+4 (\zeta - 2 \eta U)\wp''\right\rbrace  
  \frac{\D U}{\D \Omega}
\right.\nonumber \\ & & {\hskip -9.0truemm} \phantom{1234567890123456789}\left.
+ \omega^3 
\left(2 (\zeta - 2 \eta U)^2 \wp'' + 12 (\zeta - 2 \eta U) \wp \wp' + 3 {\wp'}^2\right)
\right\rbrack.
\label{eqd2usurdx2}
\end{eqnarray}

The transformed equation of (\ref{eqPVI})
then results from the substitution 
of the expressions $x$, $u$, $\D u / \D x$, $\D^2 u / \D x^2$
respectively defined by
(\ref{eq-ux-Weierstrass-rep}),
(\ref{eqdusurdx}),
(\ref{eqd2usurdx2}),
\begin{eqnarray}
& & {\hskip -7.0truemm}
\frac{\pi^2}{(2 \omega)^{3}} \frac{\D^2 U}{\D \Omega^2}
- \theta_\infty^2 \wp'
\nonumber \\ & & {\hskip -7.0truemm} \phantom{12}
-\theta_0^2 \frac{(e_3-e_1)(e_1-e_2)\wp'}{(\wp-e_1)^2}
-\theta_1^2 \frac{(e_1-e_2)(e_2-e_3)\wp'}{(\wp-e_2)^2}
-\theta_x^2 \frac{(e_2-e_3)(e_3-e_1)\wp'}{(\wp-e_3)^2}=0.
\label{eqConvert7}
\end{eqnarray}

Finally, the addition formula 
\begin{eqnarray}
& &
\forall x_1,x_2:\
\wp(x_1+x_2)+\wp(x_1)+\wp(x_2)=\frac{1}{4}
\left(\frac{\wp'(x_1)-\wp'(x_2)}{\wp(x_1)-\wp(x_2)}\right)^2,
%\label{eqWeierstrassAddition}
\end{eqnarray}
applied to the choice $x_2=\omega_\alpha$, $\wp(x_2)=e_\alpha$ reduces to
\begin{eqnarray}
& & \forall z:\
\wp(z+\omega_\alpha)-e_\alpha=\frac{(e_\alpha-e_\beta)(e_\alpha-e_\gamma)}{\wp(z)-e_\alpha},
%\label{eqWeierstrassAddition-halfperiod}
\end{eqnarray}
whose $z$-derivatives are the last three terms of (\ref{eqConvert7}).

The usual normalization (\ref{eqPVIUX})
is achieved by taking the independent variable $X$
to be an adequate multiple of $\Omega$, cf.~Eq.~(\ref{eqdefX}).

% ============================================================================
\subsection{From elliptic to rational coordinates}
% ============================================================================
\label{AppendixElliptic3var-elliptic-to-rational}

Because of homogeneity, any function of the elliptic coordinates $(U,\Omega,T)$
can be written as the product of a function of $(u,x,t)$
by a monomial of
$\omega$ (or $(e_2-e_1)$ since $(e_2-e_1) \omega^2$ has no dimension).
For instance, Eq.~(\ref{eqdefuxt}) implies
\begin{eqnarray}
& & {\hskip -13.0truemm}
\frac{\wp(2 \omega U)         }{\displaystyle               u   -\frac{x+1 }{3}}=
\frac{\wp(2 \omega U+\omega_1)}{\displaystyle \frac{x}     {u}  -\frac{x+1 }{3}}=
\frac{\wp(2 \omega U+\omega_2)}{\displaystyle-\frac{x-1}   {u-1}-\frac{x-2 }{3}}=
\frac{\wp(2 \omega U+\omega_3)}{\displaystyle-\frac{x(x-1)}{u-x}+\frac{2x-1}{3}}
%
%\frac{\wp(2 \omega U)             }{\displaystyle               u   -\frac{x+1 }{3}}=
%\frac{\wp(2 \omega U+\omega_1)-e_1}{\displaystyle \frac{x}     {u}}=
%\frac{\wp(2 \omega U+\omega_2)-e_2}{\displaystyle-\frac{x-1}   {u-1}}=
%\frac{\wp(2 \omega U+\omega_3)-e_3}{\displaystyle-\frac{x(x-1)}{u-x}}
\nonumber\\ & & {\hskip -13.0truemm}
=
\frac{3 e_1}{-(x+1)}=
\frac{3 e_2}{-(x-2)}=
\frac{3 e_3}{2x-1  }=
\frac{e_2-e_1}{1}=
\frac{e_3-e_2}{x-1}=
\frac{e_1-e_3}{-x},
\label{eqrejtox}
\end{eqnarray}
and similarly by changing $(u,U)$ to $(t,T)$.

The expressions of $\D U$, $\D T$, $\D \Omega$ 
in terms of $\D u$, $\D t$, $\D x$ %$u$, $t$, $x$, 
result from the differentials (\ref{eqdxdu})
and their transform under $(u,U) \to (t,T)$,
\begin{eqnarray}
& & {\hskip -9.0truemm}
\left\lbrace 
\begin{array}{ll}
\displaystyle{
\D U=
\frac{(e_2-e_1)} {2 \omega \wp'(2\omega U)}\D u
- \frac
{(e_2-e_1)\left[(\zeta(2\omega U) - 2 \eta U)\wp'(2\omega U)+2(\wp(2\omega U)-e_1)(\wp(2\omega U)-e_2)\right]} 
{2 (e_3-e_1)(e_3-e_2) \omega \wp'(2\omega U)} \D x,
}\\ \displaystyle{
}\\ \displaystyle{
\D T=
\frac{(e_2-e_1)} {2 \omega \wp'(2\omega T)}\D t
- \frac
{(e_2-e_1)\left[(\zeta(2\omega T) - 2 \eta T)\wp'(2\omega T)+2(\wp(2\omega T)-e_1)(\wp(2\omega T)-e_2)\right]} 
{2 (e_3-e_1)(e_3-e_2) \omega \wp'(2\omega T)} \D x,
}\\ \displaystyle{
}\\ \displaystyle{
\D \Omega=-\frac{\pi^2 (e_2-e_1)}{4 \omega^2 (e_3-e_1)(e_3-e_2)} \D x.
}
\end{array}
\right. 
\label{eqdUdTdOmega-UXT}
\end{eqnarray}
After elimination of 
$e_3$, $\wp(2 \omega U)$, $\wp(2 \omega T)$, 
$\wp'(2 \omega U)$, $\wp'(2 \omega T)$
with the definitions (\ref{eqdefuxt}), 
the one-form $\pi^2 \D U \D T / \D \Omega$, defined by
\begin{eqnarray}
& & {\hskip -13.0truemm}
\left\lbrace 
\begin{array}{ll}
\displaystyle{
\frac{\D U}{\D \Omega}=
-\frac{\omega(e_2-e_1)^{1/2}}{\pi^2\sqrt{u(u-1)(u-x)}} 
\left[x(x-1)\frac{\D u}{\D x}-u(u-1)    -\sqrt{u(u-1)(u-x)}W(u,x)\right],
}\\ \displaystyle{
\D T= {\hskip 6.0truemm}
\frac{(e_2-e_1)^{-1/2}} {4 \omega \sqrt{t(t-1)(t-x)}}
\left[x(x-1)      \D t      -t(t-1)\D x-\sqrt{t(t-1)(t-x)}W(t,x) \D x\right]
 \frac{1}{x(x-1)},
}\\ \displaystyle{
\D \Omega= -\frac{\pi^2}{4 \omega^2 (e_2-e_1) x (x-1)} \D x,
}
\end{array}
\right. 
\label{eqdUdTdOmega-uxt}
\end{eqnarray}
%\label{eqdT}
%\label{eqdOmega}
depends algebraically on $u$, $t$, $x$, $\D u$, $\D t$, $\D x$
and on a dimensionless function $W$ defined by
\begin{eqnarray}
& & {\hskip -10.0truemm}
W(u,x)=\frac{\zeta(2\omega U) - 2 \eta U}{\sqrt{e_2-e_1}},\
W(t,x)=\frac{\zeta(2\omega T) - 2 \eta T}{\sqrt{e_2-e_1}},
\label{eqWdef}
\end{eqnarray} 
which obeys a closed differential system
with coefficients depending only on $u$, $t$, $x$, $\D u$, $\D t$, $\D x$,
\begin{eqnarray}
& & {\hskip -10.0truemm}
%eqdvar0=2*dvar0*sqrt(fprod(u,x))*x*(x-1)
%       -dx*var0*sqrt(fprod(u,x))*var3
%       -(u*(u-1)*dx-x*(x-1)*du)*(var1+u)
%       +dx*sqrt(fprod(u,x))**2;
%eqdvar3=-2*dvar3*x*(x-1)+dx*(var3**2+x*(x-1));
%eqdvar1=-2*dvar1*x*(x-1)+dx*(var3**2-x*(x-1));
% var3-var1-x=0
\D W(z,x)=\frac{W(z,x)}{2 x (x-1)} V_3(x) \D x
-\frac{\sqrt{z(z-1)(z-x)}}{2 x (x-1)} \D x
+\left[z-x+V_3(x)\right]\frac{z(z-1) \D x -x(x-1)\D z}{2 x(x-1)\sqrt{z(z-1)(z-x)}},
\nonumber \\ & & {\hskip -10.0truemm}
\D V_3(x)= \left[\frac{1}{2} + \frac{V_3^2(x)}{2 x(x-1)}\right] \D x,\
 V_3(x)=\frac{\eta+e_3 \omega}{(e_2-e_1) \omega}\cdot %j=1,3,
\label{eqdWzxV1V3}
\end{eqnarray}

This system integrates with the complete elliptic integrals,
\begin{eqnarray}
& & {\hskip -13.0truemm}
\left\lbrace 
\begin{array}{ll}
\displaystyle{
V_3=-2 x(x-1) (\log \psi)',\ \psi=c_1 K(\sqrt{x})+c_2 K_{\rm C}(\sqrt{x}),\
}\\ \displaystyle{
K(k)        =\int_0^1 \frac{\D \lambda}{\sqrt{(1-\lambda^2)(1- k^2    \lambda^2)}},\
K_{\rm C}(k)=\int_0^1 \frac{\D \lambda}{\sqrt{(1-\lambda^2)(1-(k^2-1) \lambda^2)}}.
}
\end{array}
\right. 
\end{eqnarray}

\vfill\eject

% ============================================================================
\section{The solution of Bonnet to his problem}
\label{sectionBonnet_problem_and_solution}
\label{sectionLasolution}
% ============================================================================

\textbf{Problem} \cite[\S 11 pp.~72--73]{Bonnet1867}.
Given a surface in $\mathbb{R}^3$, 
to find all surfaces which are applicable\footnote{This means that both
surfaces have the same first fundamental form.} on that surface 
and possess the same two principal radii of curvature.   
\smallskip 

\textbf{Solution}. 
Using conformal coordinates, % cf note en bas de la page 32, Bonnet 1867,
Bonnet gave a complete solution to his problem in $\mathbb{R}^3$ \cite[\S 11--12 pp 72--92]{Bonnet1867}.
What we present here is the (easy to perform, see \cite{ChenLi1997}) 
extrapolation to $\Rcubec$ of his solution,
using the method of Bonnet
and the usual notation for the Gauss and Codazzi equations.
%Table \ref{Table-Bonnet-problem} (?) displays
%the correspondence of notation with Bonnet \cite[\S 11 pp.~72--73]{Bonnet1867}.
\smallskip 

Since a surface in $\Rcubec$ is characterized by ($\metric$, $H$, $Q$, $\barQ$),
the problem is equivalent to:
given a solution ($\metric$, $H$, $\mod{Q}^2$) 
of the Gauss-Codazzi equations (\ref{eqGaussCodazziR3c}),
to determine all the values of $e^{i \omega}=Q/\modQ=\modQ/\barQ$.
\smallskip 

By elimination of $\modQ$ between the two Codazzi equations
(\ref{eqGaussCodazziR3c})${}_{2,3}$,
the variable $e^{i \omega}$ obeys a second degree algebraic equation
whose coefficients only depend on $\metric$, $H$, $\modQ$ \cite[\S 11 p.~75]{Bonnet1867}
\begin{eqnarray}
& & {\hskip -10.0truemm}
%A e^{i \omega} + B e^{-i \omega} - 2 C=0,\ %\hbox{ p 75 l 4=p 86 Eq (55)} 
 H_\barz (\log\alpha)_\barz e^{ i \omega}
+H_z     (\log\beta )_z     e^{-i \omega}
-4 e^{-\metric} \modQ \metric_{z\barz} + e^\metric \modQ^{-2} H_z H_\barz=0,
\\ & & {\hskip -10.0truemm}
\alpha=e^\metric \modQ^{-2} H_\barz,\
\beta =e^\metric \modQ^{-2} H_z.
\end{eqnarray} 
Therefore the discussion splits into three cases
$(H_\barz,H_z)=$ $(0,0)$, $(0,\not=0)$, $(\not=0,\not=0)$,
the third case splits into two cases
$(\alpha_\barz,\beta_z)=$ $(0,0)$, $(\not=0,\not=0)$,
and finally the case
$(H_\barz,H_z,\alpha_\barz,\beta_z)=(\not=0,\not=0,0,0)$
splits into two cases.

Totally, there are five solutions,
summarized in Table \ref{Table-Bonnet-problem}.
%\vfill\eject
% ============================================================== Table
\tabcolsep=1.5truemm
\tabcolsep=0.5truemm

\vspace{20pt}
\begin{table}[h] % [p]
\caption[Bonnet problem]{
         The different types of analytic surfaces 
				which solve the Bonnet problem \cite[\S 11 and 12]{Bonnet1867}. 
The page numbers refer to Bonnet \cite{Bonnet1867}.	
CMC is short for constant mean curvature surfaces.	
\hfill\break
}
\vspace{0.2truecm}
\begin{center}
\begin{tabular}{| r | l | l | l | l | l |}
\hline % \hline % ********************************************************
& Characterization                           &Applicable surfaces& Comment & Pages \\ \hline \hline
% -----------------------------------------------------------------
1& $H_\barz=0,H_z=0$                         & CMC (two arb f + one PDE)  & sine-Gordon or Liouville  & 76--78
 \\ \hline 
% -----------------------------------------------------------------
2& $H_\barz=0,H_z\not=0$                     & one cone (two arb f)       & not real \cite{Bonnet1867}& 78--81
 \\ \hline 
% -----------------------------------------------------------------
3&${H_\barz H_z\not=0\atop h'+h^2-c^2=0}$    & Dual to minimal (two arb f)& not real \cite{Cartan1942}& 82--84
 \\ \hline 
% -----------------------------------------------------------------
4&${H_\barz H_z\not=0\atop h'+h^2-c^2\not=0}$& Bonnet surfaces (6-param)  & Painlev\'e VI             & 84--85 
 \\ \hline 
% -----------------------------------------------------------------
5& $\alpha_\barz \beta_z \not=0$             & One surface (Bonnet pair)  & 3 PDEs p 90               & 85--92
 \\ \hline 
% -----------------------------------------------------------------
\end{tabular}
\end{center}
\label{Table-Bonnet-problem}
\end{table}

These five types of applicable surfaces sharing the same first fundamental form
and mean curvature were obtained by Bonnet as follows.
\begin{enumerate}
	\item % ==================================================== 
Type 1 (constant mean curvature surfaces).	
Defined by
\begin{eqnarray}
& &
H_z=H_\barz=0,
\end{eqnarray}
this solution is characterized by 
\begin{eqnarray}
& &
\left\lbrace 
\begin{array}{ll}
\displaystyle{
\metric_{z \barz} +\frac{h^2-c^2}{2} e^{\metric} - 2 g_1^2(z) g_2^2(\barz) e^{-\metric}=0,
}\\ \displaystyle{
H=h,\
%(\log\modQ)_{z \barz}=0,\ 
%\modQ=g_1(z) g_2(\barz),\ e^{i \omega}=K g_1(z)/g_2(\barz), % add this K in .amp
%Q=\mu g_1^2(z),\ \barQ=\mu^{-1} g_2^2(\barz), % One can choose $\mu=1$
Q=g_1^2(z),\ \barQ=g_2^2(\barz),
}
\end{array}
\right.
\label{eqBonnet-type1} 
\end{eqnarray} 
in which $h$ is the constant mean curvature
and $g_1$, $g_2$ are two nonzero integration functions.

If $h^2=c^2$, the Liouville equation for $e^\metric$ integrates as
\begin{eqnarray}
& &
h^2=c^2:\ e^\metric=-(g_1(z) g_2(\barz))^2\frac{(g_3(z)+g_4(\barz))^2}{g_3'(z) g_4'(\barz)},
\label{eqBonnet-type1CMC} 
\end{eqnarray}
with $g_3$ and $g_4$ arbitrary,
and the relations (\ref{eqBonnet-type1}) and (\ref{eqBonnet-type1CMC}) 
are identical to the so-called Weierstrass representation of minimal surfaces
(Weierstrass 1863),
\begin{eqnarray}
& &
%H =(function(z,zb) c),
%Q =(function(z,zb) -feta(z)**2*deriv(psi(z),z)) in answer;
%Qc=(function(z,zb) -fetab(zb)**2*deriv(psib(zb),zb)) in answer;
%U =(function(z,zb) (feta(z)*fetab(zb)*(1+psi(z)*psib(zb)))**2) in answer\
H=c,\
Q=-\eta^2(z) \psi'(z),\
e^\metric=\left(1+\mod{\psi}^2\right)^2 \mod{\eta}^4,
\label{eqKenmotsu-rep}
\end{eqnarray}
with the correspondence
\begin{eqnarray}
& &
g_1^2=-\eta^2 \psi',\
g_2^2=-\bar\eta^2 \bar\psi',\
\frac{g_3}{\psi}=g_4 \bar\psi=\hbox{ arbitrary constant}.
%g_3=a \psi,\
%g_4=a / \bar\psi.
\end{eqnarray}
 
If $h^2\not=c^2$, 
a conformal transformation (\ref{eqConformal})
with $G_1'=\lambda g_1$, $G_2'=\lambda^{-1}g_2$
and $\lambda$ constant
maps the PDE for $\metric$ to 
the sine-Gordon equation
\begin{eqnarray}
& &
\metric_{Z \barZ} +\frac{h^2-c^2}{2} e^{\metric} - 2 e^{-\metric}=0,
\label{eqBonnet-type1SG} 
\end{eqnarray} 
and the reduced moving frame equations depend on $\lambda$
which is then a spectral parameter.
		
	\item % ==================================================== 
Type 2 (single complex cone).	
Defined by
\begin{eqnarray}
& &
H_\barz=0,\ (e^\metric \modQ^{-2} H_z)_\barz=0,\ \D H\not=0,
\end{eqnarray}
this surface is characterized by 
\begin{eqnarray}
& &
\left\lbrace 
\begin{array}{ll}
\displaystyle{
e^\metric=2 i \frac{g_3'(z) g_4'(\barz)}{c \cos g_3(z)},\
H=c \sin g_3(z),\
}\\ \displaystyle{
%\modQ=g_3'(z) g_4'(\barz),\ e^{i \omega}=i \frac{g_3'(z) g_4(\barz)}{g_4'(\barz)},
    Q=  i (g_3'(z))^2   g_4(\barz),\ 
\barQ= -i (g_4'(z))^2 / g_4(\barz),
}
\end{array}
\right.
\label{eqBonnet-type2} 
\end{eqnarray}
and it is not real \cite[p.~81]{Bonnet1867}.

	\item % ==================================================== 
Types 3 and 4.	
The defining relations
\begin{eqnarray}
& &
e^\metric \modQ^{-2} H_\barz=g_1(z)\not=0,\
e^\metric \modQ^{-2} H_z    =g_2(\barz)\not=0,
\label{eqBonnet-types34-def}
\end{eqnarray}
imply that
$H$ and $e^\metric \modQ^{-2}$
only depend on one variable,
whose differential is $(1/2) (g_1 \D z + g_2 \D \barz)$.
After the conformal transformation (\ref{eqConformal})
with $G_1'=2/g_1$, $G_2'=2/g_2$ followed by the elimination of $\metric$,
the variables $H=h(\xi)$ and $Q(z,\barz)$ obey the coupled system
\begin{eqnarray}
& & {\hskip -10.0truemm}
\left\lbrace 
\begin{array}{ll}
\displaystyle{
(\log h')'' + 2 h'
-8 \modQ^2 \left(\frac{h' + h^2-c^2}{h'}\right)=0,\
\xi=\frac{z+\barz}{2},\
}\\ \displaystyle{
(\log \modQ)_{z \barz} - \modQ^2=0,
}
\end{array}
\right.
\label{eqBonnet-types34} 
\end{eqnarray} 
therefore two subcases arise,
depending on whether $h$ obeys or not the Riccati ODE defined by
canceling the coefficient of $\modQ^2$ in (\ref{eqBonnet-types34})${}_1$. 

In both cases, the remaining equations are
\begin{eqnarray}
& &
%\left(\modQ e^{ i \omega}\right)_\barz=\frac{1}{2} \modQ^2,\
%\left(\modQ e^{-i \omega}\right)_z    =\frac{1}{2} \modQ^2.
e^\metric=4 \frac{\modQ^2}{h'(\xi)},\
Q_\barz=\barQ_z    =\modQ^2.
\label{eqBonnet-types34-Codazzi} 
\end{eqnarray} 
			
	\item % ==================================================== 
	Type 3 (Surfaces dual to minimal surfaces).
	If $h$ obeys the Riccati ODE, the general solution 
	\begin{eqnarray}
& & {\hskip -10.0truemm}
\left\lbrace 
\begin{array}{ll}
\displaystyle{
e^\metric=4\frac{h_1'(z) h_2'(\barz)}
  {(h_1(z)+h_2(\barz))^2 c^2 \cosh^2 c \Re(z-z_0)},\
H= c \tanh c \Re(z-z_0),\
}\\ \displaystyle{
    Q=-\frac{h_1'(z)}{h_1(z)+h_2(\barz)},\
\barQ=-\frac{h_2'(\barz)}{h_1(z)+h_2(\barz)},\
}
\end{array}
\right.
\label{eqBonnet-type3} 
\end{eqnarray} 
depends on one arbitrary constant $\Re(z_0)$ 
and two arbitrary functions of one variable.
%i.e.~the same degree of arbitrariness than CMC surfaces. (dixit Cartan)
These analytic surfaces are not real \cite[p.~57]{Cartan1942}, and
at least when $c$ is zero
there exists a conformal transformation \cite{KVoss1993}, \cite[Remark 4.3.1 p.~68]{BE2000}
mapping them to minimal surfaces in $\mathbb{R}^3$.

	\item % ==================================================== 
Type 4 (Bonnet surfaces).
They are characterized by
\begin{eqnarray}
& & 
e^\metric \modQ^{-2} H_\barz=g_1(z)\not=0,\
e^\metric \modQ^{-2} H_z    =g_2(\barz)\not=0,\
\nonumber\\ & &
\frac{2 \D H}{g_1 \D z + g_2 \D \barz}+H^2-c^2 \not=0.
\label{Type4Characterization}
\end{eqnarray}	
Since $\modQ^2$ is defined by (\ref{eqBonnet-types34})${}_1$,
it only depends on $\Re(z)$
and the equation (\ref{eqBonnet-types34})${}_2$ 
is an ODE for $\modQ$ which integrates as
\begin{eqnarray} 
\modQ= \varepsilon \frac{2 \cz}{\sinh 4 \cz \Re(    z-    z_0)},\ \varepsilon^2=1,\ 
\end{eqnarray} 
in which $\Re(z_0)$ and $\cz$ are arbitrary,
while the Codazzi equations (\ref{eqBonnet-types34-Codazzi}) 
integrate as
\begin{eqnarray}
& & {\hskip -16.0truemm}
\left\lbrace 
\begin{array}{ll}
\displaystyle{
    Q=2 \cz \coth 2 \cz(    z-    z_0) - 2 \cz \coth 4 \cz \Re(    z-    z_0)
      =\frac{\sinh 2 \cz (\barz-\barz_0)}{\sinh 2 \cz (    z-    z_0)}\frac{2 \cz}{\sinh 4 \cz \Re(    z-    z_0)},\
}\\ \displaystyle{
\barQ=2 \cz \coth 2 \cz(\barz-\barz_0) - 2 \cz \coth 4 \cz \Re(    z-    z_0)
     =\frac{\sinh 2 \cz (z-z_0)}{\sinh 2 \cz (\barz-\barz_0)}\frac{2 \cz}{\sinh 4 \cz \Re(    z-    z_0)},
}
\end{array}
\right.
\label{eq-Bonnet-type4-Q}
\end{eqnarray} 
with $z_0$ and $\barz_0$ arbitrary.
As a consequence, $\omega=\arg Q$ is characterized by the nice relation 
\cite[Eq.~(53) p.~85]{Bonnet1867},
\begin{eqnarray}
& &
 \tan \frac{\omega}{2}
=i \frac{\tanh \cz (z-z_0-\varepsilon(\barz-\barz_0))}{\tanh \cz (z-z_0+\varepsilon(\barz-\barz_0))},\
\varepsilon^2=1. 
\end{eqnarray} 
As to $H(z,\barz)=h(\xi)$, with $\xi=\Re(z)$,
it obeys the third order ODE \cite[Eq.~(52) p.~84]{Bonnet1867},
\begin{eqnarray}
& & {\hskip -16.0truemm}
(\log h')'' + 2 h'
-2\left(\frac{4 \cz}{\sinh 4 \cz (\xi-\xi_0)}\right)^2
  \left(\frac{h'+h^2-c^2}{h'}\right)=0,
\label{eq-Bonnet-type4-ODEh}
\end{eqnarray}
which admits the first integral
(\cite[p.~48]{H1897} for $c=0$, \cite{BE2000} for arbitrary $c$) 
\begin{eqnarray}
& & {\hskip -15.0 truemm}
K= 
\left(\frac{h''}{h'} + 8 \cz \coth 4 \cz (\xi-\xi_0)\right)^2
\nonumber \\ & & {\hskip -15.0 truemm} \phantom{1234}
+8 \left[
             \left(\frac{4 \cz}{\sinh 4 \cz (\xi-\xi_0)}\right)^2 \frac{h^2 -c^2}{h'}
             + h'
             + 8 \cz \coth 4 \cz (\xi-\xi_0) h\right].
\label{eqHazzi-ODE2h} 
\end{eqnarray}
The general solution \cite{BE1998} of this ODE 
(which Bonnet could not obtain for obvious chronological reasons)
is a Hamiltonian of 
either a codimension-two   $\PVI$ ($\cz\not=0$) 
or     a codimension-three $\PV$  ($\cz    =0$).
%equivalent to           a codimension-? $\PIII$.
This defines a family of analytic surfaces,
called Bonnet surfaces,
which,
in addition to the fixed parameter $c$,
depend on six arbitrary movable constants
(the two origins of $z$ and $\barz$,
 the first integrals $\cz$ and $K$,
 and the two constants of integration of the ODE (\ref{eqHazzi-ODE2h})).
Their main property is to be applicable on a surface of revolution 
but to never be a surface of revolution \cite[Eq.~(53) p.~85]{Bonnet1867}.
The real surfaces defined by these analytic surfaces
have been determined by \'E.~Cartan \cite{Cartan1942},
they require $\cz^2$ to be real and
consist of three disjoint classes denoted $A$, $B$, $C$
corresponding respectively to $\cz^2$ negative, positive, zero.

\textit{Remark} 1.
Bonnet surfaces are characterized by the local condition
\begin{eqnarray}
& &
\modQ^2 (\log \modQ)_{z \barz} - g_1(z) g_2(\barz)/4=0, 
\label{eqLiouville-Q-Bonnet-surface}
\end{eqnarray}
i.e., after elimination of $g_1$ and $g_2$, by the global conditions \cite{ChenLi1997} 
\begin{eqnarray}
& & {\hskip -16.0truemm}
(\log     Q)_{z\barz}-(\log     Q)_{\barz}(\log \barQ)_{    z}=0,\ 
(\log \barQ)_{z\barz}-(\log \barQ)_{    z}(\log     Q)_{\barz}=0. 
\label{eqBonnet-surface}
\end{eqnarray}

\textit{Remark} 2.
Since a Liouville PDE such as (\ref{eqLiouville-Q-Bonnet-surface})
is equivalent to a d'Alembert PDE $\varphi_{z \barz}=0$ (i.e.~$\varphi$ harmonic),
many geometers like to characterize Bonnet surfaces by the condition
that some function (e.g.~$Q-\barQ$ in Eq.~(\ref{eq-Bonnet-type4-Q}),
or $1/Q$ in \cite{H1897}) be harmonic,
but one should keep in mind that such a condition is only local.

%The third order ODE as written by Bonnet is \cite[Eq.~(52) p.~84]{Bonnet1867},
%\frac{1}{2}\frac{\D^2 \log \varphi'}{\D u^2}	
%- \frac{\alpha^2}{\sin^2(\alpha u)} \frac{\varphi'+\varphi^2}{\varphi'}  
%+ \varphi'=0,\ 

	\item % ==================================================== 
Type 5 (Bonnet pairs).	
They are characterized by
\begin{eqnarray}
& & {\hskip -23.0truemm}
(e^u \modQ^{-2} H_\barz)_\barz\not=0,\
(e^u \modQ^{-2} H_z    )_    z\not=0.
\end{eqnarray}	
Since there are exactly two applicable surfaces,
the proof of Bonnet can be simplified as follows \cite[\S 4.8.1]{BE2000}.
Denoting $Q_j,j=1,2$ the two solutions,
the difference of the two sets of Gauss-Codazzi equations,
\begin{eqnarray}
& &
Q_1 \barQ_1-Q_2 \barQ_2=0,\
(Q_1-Q_2)_\barz=0,\
(\barQ_1-\barQ_2)_z=0,\
\end{eqnarray}
integrates as
\begin{eqnarray}
& & {\hskip -12.0truemm}
    Q_1= \frac{\fnewq}{g_2} + g_1,\
    Q_2= \frac{\fnewq}{g_2} - g_1,\
\barQ_1=-\frac{\fnewq}{g_1} + g_2,\
\barQ_2=-\frac{\fnewq}{g_1} - g_2,\
\end{eqnarray}
in which $\fnewq(z,\barz)$, $g_1(z)$ and $g_2(\barz)$ are integration functions.
Then, to the half-sum of the two sets of Gauss-Codazzi equations
\begin{eqnarray}
& &
\left\lbrace 
\begin{array}{ll}
\displaystyle{
\metric_{z \barz} +\frac{H^2-c^2}{2} e^{\metric} + 2 \left(\frac{\fnewq^2}{g_1 g_2} - g_1 g_2\right)  e^{-\metric}=0, 
}\\ \displaystyle{
\left(\frac{\fnewq}{ g_2}\right)_\barz -\frac{1}{2} e^{\metric} H_    z=0,\
\left(\frac{\fnewq}{-g_1}\right)_    z -\frac{1}{2} e^{\metric} H_\barz=0,
}
\end{array}
\right.
\label{eqBonnet-type5} 
\end{eqnarray} 
one applies the conformal transformation (\ref{eqConformal}), 
completed by
\begin{eqnarray}
& & \forall G(z):\
(z,\fnewq) \to \left(G(z), \mod{G'(z)}^2 \fnewq \right).
\end{eqnarray}
The system resulting from the choice ${G'}^2=g_1$, ${\overline{G}'}^2=g_2$
\cite[p.~90]{Bonnet1867}
\begin{eqnarray}
& &
\left\lbrace 
\begin{array}{ll}
\displaystyle{
\metric_{z \barz} + \frac{1}{2} (H^2-c^2) e^\metric -2 (1-\fnewq^2) e^{-\metric}=0,
}\\ \displaystyle{
 \fnewq_\barz-\frac{1}{2} H_    z e^\metric=0,\ 
-\fnewq_    z-\frac{1}{2} H_\barz e^\metric=0,
}
\end{array}
\right.
\label{eq-Bonnet-type5-GC} 
\end{eqnarray} 
is a particular Gauss-Codazzi system in $\Rcubec$, with
\begin{eqnarray}
& &
Q=1+\fnewq,\ \barQ=1-\fnewq.
\end{eqnarray} 

A member $Q=Q_j$ of a Bonnet pair is characterized by the local condition
\begin{eqnarray}
& &
\frac{Q}{g_1} + \frac{\barQ}{g_2} + (-1)^j 2=0,
\end{eqnarray}
i.e., after elimination of $g_1$ and $g_2$, by the two global conditions
\begin{eqnarray}
& &
\left(\frac{
\left(\frac{\barQ_z}{Q\barQ}\right)_\barz}
{\left(\log\frac{Q}{\barQ}\right)_{z\barz}}
\right)_\barz=0 
\hbox{ and c.c.}.
\end{eqnarray}

This terminates the solution given by Bonnet to his problem.

\end{enumerate}

The five types of surfaces which solve the Bonnet problem
can also be characterized by 
the following global conditions which only involve $Q$ and $\barQ$, 
\begin{eqnarray}
& & {\hskip -16.0truemm}
\Rcubec:\
\left\lbrace 
\begin{array}{ll}
\displaystyle{
% ================== CMC
1.\ Q_\barz=0,\ \barQ_z=0, %H_z=0,\ H_\barz=0,
}\\ \displaystyle{
% ================== Cone
2.\ (\log Q)_{z \barz}=0,\ (\log \barQ)_{z \barz}=0,
}\\ \displaystyle{
% ================== Dual to CMC
3.\ (Q_\barz/\modQ^2)_z=0,\ (\barQ_z/\modQ^2)_\barz=0,\ % local g_1(z)=2 \barQ_z/\modQ^2,\ g_2(\barz)=2 Q_\barz/\modQ^2,\ 
}\\ \displaystyle{
%3.\ ? 4 e^{-\metric} H_{z \barz} + \left(\frac{1}{R_1^2}+\frac{1}{R_2^2}\right)^2=0,\ \hbox{\cite[p.~57 line 8]{Cartan1942}}
%3.\ ? 4 e^{-\metric} H_{z \barz} + 2 H (H^2-K)=0,\ \hbox{(Willmore) \cite[Remark 4.3.1]{BE2000} }
% ================== Bonnet surfaces
4.\ (\log     Q)_{z\barz}-(\log     Q)_{\barz}(\log \barQ)_{    z}=0,\ 
    (\log \barQ)_{z\barz}-(\log \barQ)_{    z}(\log     Q)_{\barz}=0,\
}\\ \displaystyle{
% ================== Bonnet pairs
5.\ \left(\frac{
\left(\frac{\barQ_z}{Q\barQ}\right)_\barz}
{\left(\log\frac{Q}{\barQ}\right)_{z\barz}}
\right)_\barz=0 
\hbox{ and c.c.}.
}
\end{array}
\right.
\label{eqBonnet-types} 
\end{eqnarray} 

\vfill \eject

% ============================================================================
\section{Confluence to the lower Painlev\'e equations}
% ============================================================================
\label{AppendixConfluence}

Following Garnier \cite{GarnierThese},
we define the lower $\Pn$ as four-parameter equations
derived 
from $\PVI(u,x,\alpha,\beta,\gamma,\delta)$
by the classical confluence of poles.

Definition in rational coordinates \cite{GarnierThese},
\label{P6P0def}
\begin{eqnarray*}
\PVI\ : \ u'' &=&
\frac{1}{2} \left[\frac{1}{u} + \frac{1}{u-1} + \frac{1}{u-x} \right] u'^2
- \left[\frac{1}{x} + \frac{1}{x-1} + \frac{1}{u-x} \right] u'
\\ & &
+ \frac{u (u-1) (u-x)}{x^2 (x-1)^2}
  \left[\alpha + \beta \frac{x}{u^2} + \gamma \frac{x-1}{(u-1)^2}
        + \delta \frac{x (x-1)}{(u-x)^2} \right],
\\ \PV\ : \ u'' &=&
\left[\frac{1}{2 u} + \frac{1}{u-1} \right] u'^2
- \frac{u'}{x}
+ \frac{(u-1)^2}{x^2} \left[ \alpha u + \frac{\beta}{u} \right]
+ \gamma \frac{u}{x}
+ \delta \frac{u(u+1)}{u-1},
\nonumber\\
\PIII\ : \ u'' &=&
\frac{u'^2}{u} - \frac{u'}{x} + \frac{\alpha u^2 + \gamma u^3}{4 x^2}
 + \frac{\beta}{4 x}
 + \frac{\delta}{4 u},
\\ \PIV'\ : \ u'' &=&
\frac{u'^2}{2 u} + \gamma \left(\frac{3}{2} u^3 + 4 x u^2 + 2 x^2 u\right)
+ 4 \delta (u^2 + x u) - 2 \alpha u + \frac{\beta}{u},
\\ \PII'\ : \ u'' &=&
\delta (2 u^3 + x u) + \gamma (6 u^2 + x) + \beta u + \alpha,
\\ {\rm J}\ : \ u'' &=&
2 \delta u^3 + 6 \gamma u^2 + \beta u + \alpha.
\end{eqnarray*}
The added equation J (like ``Jacobi'') is the autonomous limit of $\PII'$,
which is itself the synthesis of $\PII$ and $\PI$ made by Garnier.

% ============================================================
Transformation between rational and elliptic or degenerate elliptic coordinates \cite{BB1999},
\begin{eqnarray*}
   \PVI\  : \ & & x      =\frac{              e_3-e_1}{e_2-e_1}\ccomma\
u=\frac{\wp(2 \omega U,g_2,g_3)-e_1}{e_2-e_1},\
t=\frac{\wp(2 \omega T,g_2,g_3)-e_1}{e_2-e_1},\ \
%\hbox{Eqs}.~(\ref{eqdefUXT}),\ (\ref{eqdefuxt}),
\\ \PV\   : \ & & x=e^{2 X},\ u=\coth^2 U,\ t=\coth^2 T,\ 
\\ \PIII\ : \ & & x=e^{2 X}  ,\ u=e^X e^{2 U}     ,\ t=e^X e^{2 T},
\\ \PIV'\ : \ & & x=X      ,\ u=U^2      ,\ t=T^2,
\\ \PII'\ : \ & & x=X      ,\ u=U        ,\ t=T.
\end{eqnarray*}

$\Pn$ in elliptic or degenerate elliptic coordinates \cite{BB1999},
\begin{eqnarray*}
   \PVI\  : \ & & 
\frac{\D^2 U}{\D X^2}=\frac{(2 \omega)^{3}}{\pi^2 \aX^2}
\sum_{j=\infty,0,1,x}\theta_j^2 \wp'(2\omega U+\omega_j,g_2,g_3),
\\ \PV\   : \ & & 
\frac{\D^2 U}{\D X^2}=
-2 \alpha \frac{\cosh U}{\sinh^3 U}
-2 \beta \frac{\sinh U}{\cosh^3 U}
-2 \gamma e^{2 X} \sinh (2 U)
-\frac{1}{2} \delta
 e^{4 X} \sinh (4 U),
\\ \PIII\ : \ & & \frac{\D^2 U}{\D X^2}=
 \frac{1}{2} e^   X (\alpha e^{2 U} + \beta  e^{-2 U})
+\frac{1}{2} e^{2 X}(\gamma e^{4 U} + \delta e^{-4 U}),
\\ \PIV'\ : \ & & \frac{\D^2 U}{\D X^2}=
-\alpha U +\frac{\beta}{2 U^3}+\gamma\left(\frac{3}{4}U^5+ 2 X U^3 + X^2 U \right)
+2 \delta (U^3+X U),
\\ \PII'\ : \ & & \frac{\D^2 U}{\D X^2}=\delta (2 U^3 + X U) + \gamma (6 U^2 + X) + \beta U+ \alpha.
\end{eqnarray*}

% ============================================================ LAX HOLOMORPHIC
% ============================================================================
\subsection{Matrix Lax pairs holomorphic in the four parameters}
\label{section_Lax_holomorphic}
% ============================================================================

The confluence preserves the unique zero $t=u$ of $M_{12}$
and the invertibility of $M_\infty$ 
under one nonvanishing condition.

$\PVI$ see (\ref{eqLax-PVI-codim0-unbalanced-holomorphic}).

\begin{eqnarray}
& & 
\PV
\left\lbrace
\begin{array}{ll} 
\displaystyle{
%matl5poly=-Part5polym1  /tm1(t)   -Part5polymi*um1(x)/x;
%matm5poly= Part5polym1*x/tm1(t)**2+Part5polym0*(1/t-1/tm1(t))
%                                  -Part5polymi/tm1(t);
L = -\frac{M_1}{t-1}-\frac{u-1}{x} M_\infty,\
M=-\frac{M_\infty}{t-1}+ \frac{x M_1}{(t-1)^2}+\left(\frac{1}{t}-\frac{1}{t-1}\right) M_0,\
}\\ \displaystyle{
%Part5polymi=matrix((2,2),2*a11,-4,-2*alpha+a11**2,-2*a11)/4;
M_\infty =
%\frac{1}{4}\begin{pmatrix} 2 \auu & -4\cr \auu^2 - 2 \alpha & -2 \auu \cr\end{pmatrix},\
 \frac{1}{2}\begin{pmatrix} 0      & -2\cr        -   \alpha &       0 \cr\end{pmatrix},\
}\\ \displaystyle{
%Part5polym0=(
%  matrix((2,2),1,0,0,-1)*u(x)*um1(x)*(-2*abr5-2*a11*u(x)*um1(x))
% +matrix((2,2),0,1,0,0)*4*u(x)**2*um1(x)**2
% +matrix((2,2),0,0,1,0)*(-(abr5+a11*u(x)*um1(x))**2-2*beta*um1(x)**2)
% )/(u(x)*um1(x)**2*4);
M_0=\frac{1}{4 u(u-1)^2}
\begin{pmatrix} 
 -2 u(u-1)\abrfive                   & 4 u^2(u-1)^2 \cr 
   -\abrfive^2-2 \beta (u-1)^2 & 2 u(u-1)\abrfive\cr
%u(u-1)(-2 \abrfive - 2 \auu u(u-1)) & 4 u^2(u-1)^2 \cr 
%  -(\abrfive+\auu u(u-1))^2-2 \beta (u-1)^2 & -u(u-1)(-2 \abrfive - 2 \auu u(u-1)) \cr
\end{pmatrix},\
}\\ \displaystyle{
%Part5polym1=
%  matrix((2,2),1,0,0,-1)*dr/2
% +matrix((2,2),0,0,1,0)
% *(dr*abr5+(gamma-dr)*um1(x)+a11*dr*um1(x)**2)/2/um1(x)**2;
M_1=2 d \begin{pmatrix} 1 & 0 \cr 0 & -1 \cr \end{pmatrix}
%+\frac{d \abrfive+(\gamma-d)(u-1)+\auu d (u-1)^2}{2(u-1)^2}
 +\frac{d \abrfive+(\gamma-d)(u-1)               }{2(u-1)^2}
        \begin{pmatrix} 0 & 0 \cr 1 &  0 \cr \end{pmatrix},\
}\\ \displaystyle{
\abrfive=x(u'-d u),\ 
%a=\hbox{ irrelevant arbitrary constant},\
}\\ \displaystyle{
-4 \det M_\infty=2 \alpha,\ 
-4 \det M_0=-2 \beta,\
-4 \det M_1=-2 \delta=d^2;
}
\end{array}
\right.
\label{eqLax-PV-holomorphic}
\end{eqnarray}

% ============================================================ LAX HOLOMORPHIC PIII
%$\PIII$ ($M$ is Fuchsian at $t=0$, nonFuchsian at $t=\infty$),
\begin{eqnarray}
& & 
\PIII
\left\lbrace
\begin{array}{ll} 
\displaystyle{
%matl3poly=-Part3polyl0/t     -Part3polymi*u(x)/x;
%matm3poly= Part3polyl0*x/t**2-Part3polymi+Part3polym0/t;
L = -\frac{L_0}{t}                  -\frac{u}{x} M_\infty,\
M=   \frac{x L_0}{t^2}+\frac{M_0}{t}-            M_\infty,\
M_\infty =
 \frac{1}{8}\begin{pmatrix} 0      & -4 \cr        - \gamma & 0       \cr\end{pmatrix},\
}\\ \displaystyle{
M_0=\frac{2 \abrthree       }{4 u} \begin{pmatrix} 1 & 0 \cr 0 & -1\cr \end{pmatrix}
    -\frac{u}{2}                   \begin{pmatrix} 0 & 1 \cr 0 &  0\cr \end{pmatrix}
	  +\frac{\gamma u + 2 \alpha}{8} \begin{pmatrix} 0 & 0 \cr 1 &  0\cr \end{pmatrix},
}\\ \displaystyle{
L_0=-\frac{d}{4}\begin{pmatrix} 1 & 0 \cr 0 & -1\cr \end{pmatrix}
	 +\frac{-2 d \abrthree +(\beta + 2 d) u}{4 u^2} \begin{pmatrix} 0 & 0 \cr 1 &  0\cr \end{pmatrix},
%}\\ \displaystyle{
%a=\hbox{ irrelevant arbitrary constant},\
}\\ \displaystyle{
\abrthree=x u' + \frac{d}{2} x,\ % + three terms for Riccati
%write -4*det(Part3polyl0);write 0==dr**2/4-answer;
%write -4*det(Part3polymi);write 0==let gamma=cr**2 in cr**2-/4answer;
-4 \det M_\infty=\frac{\gamma}{4},\ 
-4 \det L_0=-\frac{\delta}{2}=\frac{d^2}{4};
}
\end{array}
\right.
\label{eqLax-PIII-holomorphic}
\end{eqnarray}

% ============================================================ LAX HOLOMORPHIC PIV.
%$\PIV'$ ($M$ is Fuchsian at $t=?$, nonFuchsian at $t=?$),

\begin{eqnarray}
& & 
\PIV'
\left\lbrace
\begin{array}{ll} 
\displaystyle{
%Abr4='(u,1)(x)+c*u(x)**2+2*c*x*u(x);
%Part4polymi=matrix((2,2),-c,0,2*delt,c)/4;
%Part4polym0=matrix((2,2),-c*x,1,
%               (c/2)*Abr4-alpha+c+delt*(u(x)+2*x),c*x)/2;
%Part4polym1=matrix((2,2),Abr4,-2*u(x),Abr4**2/(2*u(x))+beta/u(x),-Abr4)/4;
%matm4poly=Part4polymi*t+Part4polym0+Part4polym1/t;
%matl4poly=2*(t+u(x))*Part4polymi+2*Part4polym0;
L= 2(t+u) M_\infty+2 M_0,\
M=   t    M_\infty+  M_0 + \frac{M_{-1}}{t},\
M_\infty=\frac{1}{4}\begin{pmatrix} -c & 0 \cr 2 \delta & c\cr \end{pmatrix},\
}\\ \displaystyle{
M_0=\frac{1}{4} \begin{pmatrix} -2c x & 2 \cr c \abrfour + 2 \delta (u+2 x)- 2 (\alpha + c) & 2 c x\cr \end{pmatrix},\
M_{-1}=\frac{1}{8 u}\begin{pmatrix} 2 u \abrfour & -4 u^2 \cr \abrfour^2 +2 \beta & -2 u\abrfour\cr \end{pmatrix},\
}\\ \displaystyle{
\abrfour=u'+c u^2 +2 c x u,\
%}\\ \displaystyle{
-4 \det M_\infty=\frac{\gamma}{4}=\frac{c^2}{4},\
-4 \det M_{-1}  =\frac{\beta}{2};
}
\end{array}
\right.
\label{eqLax-PIV-holomorphic}
\end{eqnarray}
		
% ============================================================ LAX HOLOMORPHIC PII'.	
%$\PII'$ %($M$ is Fuchsian at $t=?$, nonFuchsian at $t=?$),
\begin{eqnarray}
& & 
\PII'
\left\lbrace
\begin{array}{ll} 
\displaystyle{
%Abr2='(u,1)(x)+dr*(u(x)**2+x/2);
%Part2polymi=matrix((2,2),-dr,0,2*gamma,dr);
%Part2polym1=matrix((2,2),0,2,dr*Abr2+beta/2+2*gamma*u(x),0);
%Part2polym0=matrix((2,2),1,0,0,-1)*('(u,1)(x)+dr*u(x)**2)
%           +matrix((2,2),0,1,0,0)*(-2*u(x))
%           +matrix((2,2),0,0,1,0)
% *(dr*(u(x)*Abr2+1/2)+alpha+beta*u(x)/2+gamma*(2*u(x)**2+x));
%matl2poly=(t+u(x))*Part2polymi/2+(1/2)*Part2polym1;
%matm2poly=t**2*Part2polymi+t*Part2polym1+Part2polym0;
L = \frac{t+u}{2} M_\infty + \frac{M_1}{2},\
M=            t^2 M_\infty+      t M_1 + M_0,\
}\\ \displaystyle{
M_\infty =\begin{pmatrix} -d & 0 \cr  2 \gamma & d \cr\end{pmatrix},\
%}\\ \displaystyle{
M_1=\begin{pmatrix} 0 & 2 \cr d \abrtwo + \frac{\beta}{2} + 2 \gamma u & 0\cr \end{pmatrix},\
}\\ \displaystyle{
M_0=\begin{pmatrix} u'+d u^2 & -2 u \cr 
    d(2 u \abrtwo+1/2)+\alpha + \beta u/2 +\gamma (2u^2+x) & 2 -(u'+ d u^2)\cr \end{pmatrix},\
}\\ \displaystyle{
\abrtwo=u'+d\left(u^2+\frac{x}{2}\right),\ 
%}\\ \displaystyle{
-4 \det M_\infty=4 \delta=4 d^2.
}
\end{array}
\right.
\label{eqLax-PII-holomorphic}
\end{eqnarray}

For $\PV$, the two Fuchsian singularities of $M$ are naturally put
at $t=\infty$ and $t=0$ by the confluence. 
The choice $t=0,1$ for their location made in Ref.~\cite[(C.38)]{JimboMiwaII} 
breaks the symmetry between $t$ and $u$,
resulting in distorted values of the invariants $\det M_j$.

% ============================================================================
\subsection{Matrix Lax pairs symmetric with respect to the diagonal}
% ============================================================================
% ============================================================

They are generated from (\ref{eqLax-PVI-codim0-balanced-meromorphic}) by the confluence.
Alternatively, they are
obtained from those in section \ref{section_Lax_holomorphic}
by action of the transition matrix $P$ displayed in the first line of each entry.
They have a meromorphic dependence in one of the four parameters.
{}From $\PIV'$ down, $M_\infty$ is no more diagonal and all elements are rational.

$\PVI$ see (\ref{eqLax-PVI-codim0-balanced-meromorphic}).

% ============================================================ PV meromorphic rational log deriv
%$\PV$ ($M$ is Fuchsian at $t=\infty$ and $t=0$, nonFuchsian at $t=1$)
\begin{eqnarray}
& & 
\PV
\left\lbrace
\begin{array}{ll} 
\displaystyle{
\alpha=\frac{\theta_\infty^2}{2}\not=0:\ 
P=\begin{pmatrix} 2 & 2 \cr \auu-\theta_\infty & \auu+\theta_\infty \cr \end{pmatrix}
  \begin{pmatrix} g^{-1/2} & 0\cr 0 & g^{1/2} \cr\end{pmatrix},\
%matl5mero=-Partm1/tm1(t)/x;
%matm5mero= Partm1/tm1(t)**2+Partm0*(1/t-1/tm1(t))-matrix((2,2),1,0,0,-1)*ti/2/tm1(t);
}\\ \displaystyle{
L = -\frac{M_1}{t-1},\ 
M=-\frac{M_\infty}{t-1}+ \frac{x M_1}{(t-1)^2}+\left(\frac{1}{t}-\frac{1}{t-1}\right) M_0,\
}\\ \displaystyle{
M_\infty =\frac{\theta_\infty}{2}\begin{pmatrix} 1 & 0\cr 0 & -1 \cr\end{pmatrix},\
}\\ \displaystyle{
%Partm0=(matrix((2,2),1,0,0,-1)
% *( (x*'(u,1)(x)-dr*x*u(x))**2-2*(alpha*u(x)**2-beta)*um1(x)**2)
% +matrix((2,2),0,1,0,0)
% *( (x*'(u,1)(x)-dr*x*u(x)-ti*u(x)*um1(x))**2+2*beta*um1(x)**2)*fg(x)
% +matrix((2,2),0,0,1,0)
% *(-(x*'(u,1)(x)-dr*x*u(x)+ti*u(x)*um1(x))**2-2*beta*um1(x)**2)/fg(x)
%)/(u(x)*um1(x)**2*ti*4);
%M_{0,11}=\frac{1}{N}\left[\ \ \left(\abrfive                     \right)^2-2(\alpha u^2-\beta)(u-1)^2\right],\
%M_{0,12}=\frac{1}{N}\left[\ \ \left(\abrfive-\theta_\infty u(u-1)\right)^2+2 \beta(u-1)^2\right] g,\
%M_{0,21}=\frac{1}{N}\left[   -\left(\abrfive+\theta_\infty u(u-1)\right)^2-2 \beta(u-1)^2\right] g^{-1},\
%}\\ \displaystyle{
M_0=\frac{\abrfive^2+2 \beta u^2 (u-1)^2}{4 \theta_\infty u(u-1)^2}
   \begin{pmatrix} 1 & g\cr -1/g & -1 \cr\end{pmatrix}
%\right.}\\ \displaystyle{\left.\phantom{123}
	-\frac{\abrfive}{2(u-1)} \begin{pmatrix} 0 & g\cr  1/g & 0 \cr\end{pmatrix}
}\\ \displaystyle{\phantom{123}
	+\frac{\theta_\infty u}{4} \begin{pmatrix} -1 & g\cr -1/g & 1 \cr\end{pmatrix},
}\\ \displaystyle{
%
%Partm1=(matrix((2,2),1,0,0,-1)*
% (-dr*x*'(u,1)(x)-(gamma-dr)*um1(x)+dr**2*x*u(x))
% +matrix((2,2),0,1,0,0)*
% (-dr*x*'(u,1)(x)-(gamma-dr)*um1(x)+dr**2*x*u(x)+dr*ti*um1(x)**2)*fg(x)
% +matrix((2,2),0,0,1,0)*
% ( dr*x*'(u,1)(x)+(gamma-dr)*um1(x)-dr**2*x*u(x)+dr*ti*um1(x)**2)/fg(x)
%)/um1(x)**2/ti/2;
%M_{1,11}= \frac{2 u}{N}\left[   -d x u'-(\gamma-d)(u-1)+d^2 x u\right],\
%M_{1,12}= \frac{2 u}{N}\left[   -d x u'-(\gamma-d)(u-1)+d^2 x u +d \theta_\infty (u-1)^2 \right] g,\
%M_{1,21}= \frac{2 u}{N}\left[\ \ d x u'+(\gamma-d)(u-1)-d^2 x u +d \theta_\infty (u-1)^2 \right] g^{-1},\
M_1=\frac{1}{2 \theta_\infty (u-1)^2}\left[-d \abrfive-(\gamma-d)(u-1)\right]
   \begin{pmatrix} 1 & -g\cr 1/g & -1 \cr\end{pmatrix}
	+ \frac{d}{2}
   \begin{pmatrix} 0 &  g\cr 1/g & 0 \cr\end{pmatrix},
}\\ \displaystyle{
\abrfive=x(u'-d u),\ 
\frac{g'}{g}= \theta_\infty\frac{u-1}{x},\
\delta=-\frac{d^2}{2},\
% N=4 \theta_\infty u(u-1)^2,\
}\\ \displaystyle{
-4 \det M_\infty=2 \alpha=\theta_\infty^2,\ 
-4 \det M_0=-2 \beta,\
-4 \det M_1=d^2;
}
\end{array}
\right.
\label{eqLax-PV-meromorphic}
\end{eqnarray}

% ============================================================ PIII meromorphic rational log deriv
%$\PIII$ ($M$ is Fuchsian at $t=0$, nonFuchsian at $t=\infty$),
\begin{eqnarray}
& & 
\PIII
\left\lbrace
\begin{array}{ll} 
\displaystyle{
\gamma=c^2\not=0:\ 
P=\begin{pmatrix} 2 & 2\cr -c & c \cr\end{pmatrix} 
  \begin{pmatrix} g^{-1/2} & 0\cr 0 & g^{1/2} \cr\end{pmatrix},\
}\\ \displaystyle{
%matl3mero=-Part3l0/t;
%matm3mero= Part3l0*x/t**2+Part3m0/t-Part3mi;
L = -\frac{L_0}{t},\ 
M=\frac{x L_0}{t^2}+\frac{M_0}{t}-M_\infty,\
%}\\ \displaystyle{
%Part3mi=matrix((2,2),1,0,0,-1)*c/4;
M_\infty =\frac{c}{4}\begin{pmatrix} 1 & 0\cr 0 & -1 \cr\end{pmatrix},\
}\\ \displaystyle{
L_0=\frac{2 d \abrthree -(\beta + 2 d) u}{4 c u^2}
   \begin{pmatrix} 1 & g\cr -1/g & -1 \cr\end{pmatrix}
	-\frac{d}{4}
   \begin{pmatrix} 0 & g\cr  1/g &  0 \cr\end{pmatrix},
}\\ \displaystyle{
M_0=\frac{\abrthree}{2 u} \begin{pmatrix} 0 & g\cr  1/g &  0 \cr\end{pmatrix}
   -\frac{\alpha}{4 c}    \begin{pmatrix} 1 & g\cr -1/g & -1 \cr\end{pmatrix}
   -\frac{c u}{4}         \begin{pmatrix} 0 & g\cr -1/g &  0 \cr\end{pmatrix},
}\\ \displaystyle{
\abrthree=x u' + \frac{d}{2} x,\ % + three terms for Riccati
\frac{g'}{g}= c \frac{u}{2 x},\
\gamma=c^2,\
\delta=-d^2,\
}\\ \displaystyle{
%write -4*det(Part3l0);write 0==dr**2/4-answer;
%write -4*det(Part3mi);write 0==c**2/4-answer;
-4 \det M_\infty=c^2/4,\
-4 \det L_0     =d^2/4;
}
\end{array}
\right.
\label{eqLax-PIII-meromorphic}
\end{eqnarray}

% ============================================================ PIV' meromorphic rational
%$\PIV'$ 
\begin{eqnarray}
& & 
\PIV'
\left\lbrace
\begin{array}{ll} 
\displaystyle{
\gamma=c^2\not=0:\ 
P=\begin{pmatrix} 2 & 2\cr c & -c \cr\end{pmatrix},\
L= 2 (t+u) M_\infty+2 M_0,\
M=    t    M_\infty+  M_0 + \frac{M_{-1}}{t},\
}\\ \displaystyle{
%
%Part4meromi=matrix((2,2),2*delt,2*delt-c**2,-2*delt-c**2,-2*delt)/(4*c);
%Part4merom0=matrix((2,2),1, 1,-1,-1)*(c*Abr4-2*alpha+2*delt*(u(x)+2*x)+2*c)/(4*c)
%           +matrix((2,2),0, 1, 1, 0)*(-c*x/2)
%           +matrix((2,2),1,-1, 1,-1)*c/8;
%Part4merom1=matrix((2,2),1, 1,-1,-1)*(Abr4**2+2*beta)/(8*c*u(x))
%           +matrix((2,2),0, 1, 1, 0)*Abr4/4
%           -matrix((2,2),1,-1, 1,-1)*c*u(x)/8;
%matm4mero=Part4meromi*t+Part4merom0+Part4merom1/t;
%matl4mero=2*(t+u(x))*Part4meromi+2*Part4merom0;
%write -4*det(Part4meromi);write 0==(c/2)**2-answer;
%write -4*det(Part4merom1);write 0==-beta/2-answer;
M_\infty =\frac{1}{4 c} \begin{pmatrix} 2 \delta & 2 \delta-c^2 \cr -2 \delta-c^2 & -2 \delta \cr\end{pmatrix},\
}\\ \displaystyle{
M_0 =\frac{c \abrfour + 2 \delta (u+2 x) - 2 \alpha + 2 c}{4 c} 
                   \begin{pmatrix} 1 &  1 \cr -1 & -1 \cr\end{pmatrix}
		-\frac{c x}{2} \begin{pmatrix} 0 &  1 \cr  1 &  0 \cr\end{pmatrix},\
    +\frac{c}{8}   \begin{pmatrix} 1 & -1 \cr  1 & -1 \cr\end{pmatrix}
		}\\ \displaystyle{
M_{-1} =\frac{\abrfour^2 + 2 \beta}{8 c u} 
                   \begin{pmatrix} 1 &  1 \cr -1 & -1 \cr\end{pmatrix}
+\frac{\abrfour}{4}\begin{pmatrix} 0 &  1 \cr  1 &  0 \cr\end{pmatrix}
    -\frac{c u}{8} \begin{pmatrix} 1 & -1 \cr  1 & -1 \cr\end{pmatrix},\
}\\ \displaystyle{
\abrfour=u'+c u^2 +2 c x u,\
\gamma=c^2,\
-4 \det M_\infty=\frac{c^2}{4},\
-4 \det M_{-1}  =-\frac{\beta}{2};
}
\end{array}
\right.
\label{eqLax-PIV-meromorphic}
\end{eqnarray}
												
% ============================================================ PII' meromorphic rational
%$\PII'$ %($M$ is Fuchsian at $t=?$, nonFuchsian at $t=?$),
\begin{eqnarray}
& & 
\PII'
\left\lbrace
\begin{array}{ll} 
\displaystyle{
\delta=d^2\not=0:\
P=\begin{pmatrix} 2 & 2 \cr d & -d\cr \end{pmatrix},\ 
%let rule rg2  '(fg,1)(x)=fg(x)*('(u,1)(x)+dr*u(x)**2+dr*x/2+dr/2+fr(x))
%let rule rg2a '(fg,1)(x)=0,fg(x)=1,fr(x)=-Abr2-dr/2   inactive;
L = \frac{t+u}{2} M_\infty + \frac{M_1}{2},\
M=  t^2 M_\infty+t M_1 + M_0,\
}\\ \displaystyle{
M_\infty =\frac{1}{d}\begin{pmatrix} 2 \gamma & 2 \gamma-d^2 \cr  -2 \gamma-d^2 & -2 \gamma \cr\end{pmatrix},\
}\\ \displaystyle{
M_1=-\frac{d}{2}                             \begin{pmatrix} 1 & -1 \cr  1 & -1\cr \end{pmatrix}
+\frac{2 d \abrtwo + 4 \gamma u +\beta}{2 d} \begin{pmatrix} 1 &  1 \cr -1 & -1\cr \end{pmatrix},\
}\\ \displaystyle{
M_0= -\frac{d u}{2}        \begin{pmatrix} 1 & -1 \cr  1 & -1\cr \end{pmatrix}
+\frac{2 d u \abrtwo + 4 \gamma u^2 +\beta u +2 \gamma x +2 \alpha+d}{2 d} 
                           \begin{pmatrix} 1 &  1 \cr -1 & -1\cr \end{pmatrix}
}\\ \displaystyle{\phantom{12345}
+\frac{2 \abrtwo - d x}{2} \begin{pmatrix} 0 &  1 \cr  1 &  0\cr \end{pmatrix},\
}\\ \displaystyle{
\abrtwo=u'+d\left(u^2+\frac{x}{2}\right),\
-4 \det M_\infty=4 \delta=4 d^2.
}
\end{array}
\right.
\label{eqLax-PII-meromorphic}
\end{eqnarray}

% ============================================================================
\subsection{Quantum correspondence}

\textit{First step}.
The classical Hamiltonians are generated by the confluence acting on
(\ref{eqHamVIT}).
As explained in the text,
these Hamiltonians for $\PVI$, $\PV$ and $\PIII$
are different from those in Ref.~\cite{Suleimanov1994},
they only coincide at the $\PIV'$ and $\PII'$ levels 
because of the absence of a $p$ term. 

% ============================================================================
\textit{Second step}.
One defines the scalar Lax pairs (\ref{eqLaxd}) 
of the four-parameter $\Pn$'s
by their two coefficients $(S,C)$,
see Ref.~\cite[pp.~49, 52]{GarnierThese},
reproduced in 
Ref.~\cite[p.~211]{CMBook},
%   \PVI\  &:& C=-\frac{t(t-1)(u-x)}{x(x-1)(t-u)}\ccomma\ \frac{S}{2}=(\ref{eqS})
%\\ \PV\   &:& C=-\frac{t(t-1)(u-1)}{x     (t-u)}\ccomma\ \frac{S}{2}=(\ref{eqS})
%\\ \PIV'\ &:& C=-\frac{2 t        }{      (t-u)}\ccomma\ \frac{S}{2}=(\ref{eqS})
%\\ \PIII\ &:& C=-\frac{t      u   }{x     (t-u)}\ccomma\ \frac{S}{2}=(\ref{eqS})
%\\ \PII'\ &:& C=-\frac{1          }{2     (t-u)}\ccomma\ \frac{S}{2}=(\ref{eqS})
% ============================================================================
and 
one changes the scalar wave vector from $\psi_{\rm d}$ to $\psi_{\rm h}$.

% ============================================================================
\textit{Third step}.
The quantum Hamiltonians are defined from the classical ones 
as the confluence of the quantization rule (\ref{eqQuantumHamVI}).
If one chooses the normalization constant $\aT$ adequately for each $\Pn$,
the quantization rule (\ref{eqQuantumHamVI}) is the same for every $\Pn$,
\begin{eqnarray}
& & {\hskip -15.0 truemm}
\forall \Pn,\ \forall f(q,x),\ \forall k=1,2:\
f(q,x) p^k \psi_{\rm h} \to f(t,x) \aT^{-k} \partial_t^k \psi_{\rm h}. 
\label{eqQuantum}
\end{eqnarray}

The result is as follows.

Equations (\ref{eqLaxd}) (scalar Lax pair), (\ref{eqQuantum}) (quantization) 
and $q=u$
are common to all $\Pn$.

% ============================================================ PV
\begin{eqnarray}
& & {\hskip -10.0 truemm}
\PV
\left\lbrace
\begin{array}{ll} 
\displaystyle{
V(z)=\alpha z +\frac{\beta}{2} z^2 + \gamma(2 z^3+z x)+\frac{\delta}{2}(z^4+z^2 x),
}\\ \displaystyle{
\frac{\D^2 u}{\D x^2}-\left[\frac{1}{2 u} + \frac{1}{u-1} \right] u'^2+ \frac{u'}{x}
-\frac{u(u-1)^2}{x}\frac{\partial V(u)}{\partial u}=0,\
}\\ \displaystyle{
S=-\frac{3}{2 (t-u)^2}+2 \frac{x u'+(u-1)(2 u-1)}{(u-1)t (t-1) (t-u)}
-\frac{1}{2 t u}\left(\frac{x u'}{(t-1)(u-1)}\right)^2
}\\ \displaystyle{
\phantom{12345} 
+\frac{x}{t(t-1)^2}\left(V(u)-V(t) \right)
+\frac{1}{2}\left(\frac{2 t-1}{t(t-1)}\right)^2,\
C=-\frac{t(t-1)(u-1)}{x(t-u)},\
}\\ \displaystyle{
\Hqp(q,p,x,\alpha,\beta,\gamma,\delta)=\frac{q(q-1)}{x}((q-1)\aT p^2+p)-\frac{1}{2 \aT} V(q),\
}\\ \displaystyle{
\phantom{12345} 
p=\frac{1}{2 \aT}\left(\frac{x u'}{u(u-1)^2}-\frac{1}{u-1}\right),\
}\\ \displaystyle{
g_{\rm d}-g_{\rm h}=-\aT \Hqp(q,p,x,\alpha,\beta,\gamma,\delta)+\frac{x u'-(u-1)}{2 x(u-1)},\
\psi_{\rm d}= \psi_{\rm h}(t-u)^{-1/2} (t-1),\
}\\ \displaystyle{
\left[\partial_x+g_{\rm h}(x)-\aT\Hqp(t,\partial_t,x,\alpha+1/2,\beta+1/2,\gamma,\delta)
 \right]\psi_{\rm h}=0,\ 
}\\ \displaystyle{
\left\lbrack
\partial_x +g_{\rm h}(x) -\aT \Hqp(q,p,x,\alpha,\beta,\gamma,\delta)+ C \partial_t
+\frac{(x u'-(u-1))(t-1)}{2 x(u-1)(t-u)}
\right\rbrack \psi_{\rm h}=0,\
}
\end{array}
\right.
\label{eqLax-PV-quantum}
\end{eqnarray}		
	
% ============================================================ PIII, select k_1=1
\begin{eqnarray}
& & {\hskip -10.0 truemm}
\PIII
\left\lbrace
\begin{array}{ll} 
\displaystyle{
V(z)=\frac{1}{16}\left(2 \alpha \frac{z}{x} -2 \frac{\beta}{z} + \gamma\frac{z^2}{x}-\delta \frac{x}{z^2}\right),\
\left(\frac{x \D u}{u \D x}\right)'- 2 u \frac{\partial V(u)}{\partial u}=0,\
}\\ \displaystyle{
S=-\frac{3}{2 (t-u)^2}+2 \frac{x u'+u}{t u (t-u)}-\frac{1}{2}\left(\frac{x u'}{t u}\right)^2
+\frac{2 x}{t^2}\left(V(u)-V(t) \right),\
C=-\frac{t u}{x(t-u)},\
}\\ \displaystyle{
\Hqp(q,p,x,\alpha,\beta,\gamma,\delta)=\frac{q^2}{x}\aT p^2+\frac{q}{x} p-\frac{1}{\aT} V(q)+\frac{1}{4 \aT x},\
p=\frac{1}{2 \aT}\left(\frac{x u'}{u^2}-\frac{1}{u}\right),\ 
}\\ \displaystyle{
g_{\rm d}-g_{\rm h}=-\aT \Hqp(q,p,x,\alpha,\beta,\gamma,\delta) +\frac{u'}{2 u},\
\psi_{\rm d}= \psi_{\rm h}(t-u)^{-1/2} t x^{-1/2},\
}\\ \displaystyle{
\left[\partial_x+g_{\rm h}(x)-\aT\Hqp(t,\partial_t,x,\alpha,\beta,\gamma,\delta)\right]\psi_{\rm h}=0,\ 
}\\ \displaystyle{
\left\lbrack
\partial_x +g_{\rm h}(x) -\aT \Hqp(q,p,x,\alpha,\beta,\gamma,\delta)+ C \partial_t
 +\frac{(x u'-u) t}{2 x u (t-u)}
\right\rbrack \psi_{\rm h}=0,\
}
\end{array}
\right.
\label{eqLax-PIII-quantum}
\end{eqnarray}		
	
% ============================================================ PIV'
\begin{eqnarray}
& & {\hskip -10.0 truemm}
\PIV'
\left\lbrace
\begin{array}{ll} 
\displaystyle{
V(z)=-2 \alpha z -\frac{\beta}{z} + \gamma(\frac{z^3}{2}+2 x z^2+2 x^2 z)+\delta(2 z^2+4 x z),
}\\ \displaystyle{
\frac{\D^2 u}{\D x^2}-\frac{{u'}^2}{2 u}
-u \frac{\partial V(u)}{\partial u}=0,\
}\\ \displaystyle{
S=-\frac{3}{2 (t-u)^2}+2 \frac{u'+2}{2 t (t-u)}-\frac{{u'}^2}{8 t u}+\frac{1}{2 t^2}
+\frac{1}{4 t}\left(V(u)-V(t) \right),\
C=-\frac{2 t}{t-u},\
}\\ \displaystyle{
\Hqp(q,p,x,\alpha,\beta,\gamma,\delta)=q \aT p^2-\frac{1}{4 \aT} V(q),\
p=\frac{u'}{4 \aT u},\
}\\ \displaystyle{
g_{\rm d}-g_{\rm h}=-\aT \Hqp(q,p,x,\alpha,\beta,\gamma,\delta),\
\psi_{\rm d}= \psi_{\rm h}(t-u)^{-1/2},\
}\\ \displaystyle{
\left[\partial_x+g_{\rm h}(x)
      -\aT\Hqp(t,\partial_t,x,\alpha,\beta,\gamma,\delta)\right]\psi_{\rm h}=0,\ 
}\\ \displaystyle{
\left\lbrack
\partial_x +g_{\rm h}(x)-\aT \Hqp(q,p,x,\alpha,\beta,\gamma,\delta)+ C \partial_t
+\frac{u'+2}{2 (t-u)}
\right\rbrack \psi_{\rm h}=0,\
}
\end{array}
\right.
\label{eqLax-PIV-quantum}
\end{eqnarray}		
	
% ============================================================ PII' 
\begin{eqnarray}
& & {\hskip -10.0 truemm}
\PII'
\left\lbrace
\begin{array}{ll} 
\displaystyle{
V(z)=\alpha z +\frac{\beta}{2} z^2 + \gamma(2 z^3+z x)+\frac{\delta}{2}(z^4+z^2 x),\
\frac{\D^2 u}{\D x^2}+
\frac{\partial V(u)}{\partial u}=0,\
}\\ \displaystyle{
S=-\frac{3}{2 (t-u)^2}+2 \frac{u'}{t-u}-2 {u'}^2+4 V(u)-4 V(t),\
C=-\frac{1}{2 (t-u)},\
}\\ \displaystyle{
\Hqp(q,p,x,\alpha,\beta,\gamma,\delta)=\aT \frac{p^2}{2}-\frac{1}{\aT} V(q),\
p=\frac{u'}{\aT},\
}\\ \displaystyle{
g_{\rm d}-g_{\rm h}=-\aT \Hqp(q,p,x,\alpha,\beta,\gamma,\delta),\
\psi_{\rm d}= \psi_{\rm h}(t-u)^{-1/2},\
}\\ \displaystyle{
\left[\partial_x+g_{\rm h}(x)-\aT\Hqp(t,\partial_t,x,
\alpha,\beta,\gamma,\delta)\right]\psi_{\rm h}=0,\ 
}\\ \displaystyle{
\left\lbrack
\partial_x +g_{\rm h}(x)-\aT \Hqp(q,p,x,\alpha,\beta,\gamma,\delta)+ C \partial_t
+\frac{u'}{2 (t-u)}
\right\rbrack \psi_{\rm h}=0.
}
\end{array}
\right.
\label{eqLax-PII-quantum}
\end{eqnarray}		
						
% ============================================================================
\subsection{Generalized heat equations and associated Lax pairs}
% ============================================================================

The confluence of the relations (\ref{eqLaxhuxt}) yields for each $\Pn$
a scalar Lax pair made of a generalized heat equation and a first order PDE.

If one denotes $\Hqp_{\rm n}(q,p,x,\alpha,\beta,\gamma,\delta)$
the above classical Hamiltonians,
the generalized heat equations are (we omit the $g_{\rm h}(x)$ terms),
%\begin{eqnarray}
%& & {\hskip -15.0 truemm}
%\left\lbrace 
%\begin{array}{ll}
%\displaystyle{
%%-Laxh2+x*xm1(x)*'(psih,1,0)(x,t)+(-x*xm1(x)*aT*hamTshifted+gh(x))*psih(x,t);
%\left[\partial_x-\aT\Hqp_{\rm VI}(t,\partial_t,x,\theta_{j}^2+s_j)\right]\psi_{\rm h}=0,\ %RC
%s_\infty=1,\ s_0=s_1=s_x=-1, %RC
%}\\ \displaystyle{
%%
%%-Lax5h2+x*'(psih,1,0)(x,t)+(-x*aT*ham0shifted+gh(x))*psih(x,t);
%%let si=1/2,s0=1/2,s1=0,sx=0 in answer;
%\left[\partial_x-\aT\Hqp_{\rm V}(t,\partial_t,x,\alpha+1/2,\beta+1/2,\gamma,\delta)\right]\psi_{\rm h}=0,\ %RC
%%s_\infty=s_0=\frac{1}{2},\ s_1=s_x=0,
%}\\ \displaystyle{
%%
%%-Lax3h2+x*'(psih,1,0)(x,t)+(-x*aT*(ham0shifted)+gh(x))*psih(x,t);
%%let si=0,s0=0,s1=0,sx=0 in answer;
%\left[\partial_x-\aT\Hqp_{\rm n}(t,\partial_t,x,\alpha,\beta,\gamma,\delta)\right]\psi_{\rm h}=0,\ \hbox{n=III, IV', II'}. %RC
%%s_\infty=s_0=s_1=s_x=0,
%%
%%-Lax4h2+'(psih,1,0)(x,t)+(-aT*ham0shifted+gh(x)/x)*psih(x,t);
%%let si=0,s0=2,s1=0,sx=0 in answer;
%%s_\infty=s_0=s_1=s_x=0,
%%
%%-Lax2h2+'(psih,1,0)(x,t)+(-aT*ham0shifted+gh(x)/x)*psih(x,t);
%%write 0==let si=0,s0=0,s1=0,sx=0,rab2=-aT*f(x) in answer;
%%s_\infty=s_0=s_1=s_x=-1,
%%}
%%\end{array}
%%\right.
%%\label{eqHamnHeat}  
%\end{eqnarray} 
\begin{eqnarray}
& & {\hskip -15.0 truemm}
\forall \Pn:\ 
\left[\partial_x-\aT\Hqp_{\rm n}(t,\partial_t,x,
\alpha+s_\alpha,\beta+s_\beta,\gamma+s_\gamma,\delta+s_\delta)\right]\psi_{\rm h}=0,\ 
\label{eqHamnHeat}  
\end{eqnarray} 
in which the shifts $s_*$ of the parameters are nonzero only for $\PVI$ and $\PV$,
\begin{eqnarray}
& & {\hskip -15.0 truemm}
(s_\alpha,s_\beta,s_\gamma,s_\delta)= 
\left\lbrace 
\begin{array}{ll}
\displaystyle{
(1/2,1/2,-1/2,1/2),\ \PVI % \theta_j^2+(1,-1,-1,-1)
}\\ \displaystyle{
(1/2,1/2,0,0),\ \PV
}\\ \displaystyle{
(0,0,0,0),\ \PIII, \PIV', \PII'.
}
\end{array}
\right.
\label{eqHamnshifts}  
\end{eqnarray} 

The second half of the Lax pairs is as follows,

% ============================================================ PV Lax5h1
\begin{eqnarray}
& & {\hskip -10.0 truemm}
\PV
\left\lbrace
\begin{array}{ll} 
\displaystyle{
%Lax5h1=
\left\lbrack
\partial_x - \frac{t(t-1)(u-1)}{x(t-u)} \partial_t
+\frac{x u'-u+1}{2 x(t-u)}
-\frac{x {u'}^2}{4 u(u-1)^2}
+\frac{x  u'   }{2 x(u-1)}
\right.}\\ \displaystyle{\left.\phantom{1}
+\alpha \left(\frac{u}{2 x}-\frac{1}{4 x}\right)
+\beta  \left(\frac{1}{4 x}-\frac{1}{2 x u}\right)
+\gamma \left(\frac{1}{4}-\frac{u}{2(u-1)}\right)
-\delta \frac{x u}{2(u-1)^2}
\right\rbrack \psi_{\rm h}=0,
}\\ \displaystyle{
}
\end{array}
\right.
\label{eqLax-PV-second-half}
\end{eqnarray}

% ============================================================ PIII Lax3h1
\begin{eqnarray}
& & {\hskip -10.0 truemm}
\PIII 
\begin{array}{ll} 
\displaystyle{
%Lax3h1=
\left\lbrack
\partial_x - \frac{t u}{x(t-u)} \partial_t
+\frac{x u'-u}{2 (t-u)}
-\frac{x{u'}^2}{4 u^2}
+\frac{  u'   }{2 u}-\frac{1}{2 x}
+\frac{1}{8}\left(
 \alpha \frac{u}{x}
-\frac{\beta}{u} 
+\gamma \frac{u^2}{2 x}
-\delta \frac{x}{2 u^2}
\right)
\right\rbrack \psi_{\rm h}=0,
}
\end{array}
\label{eqLax-PIII-second-half}
\end{eqnarray}

% ============================================================ PIV' Lax2h1
\begin{eqnarray}
& & {\hskip -10.0 truemm}
\PIV'
\left\lbrace
\begin{array}{ll} 
\displaystyle{
%Lax4h1=
\left\lbrack
\partial_x - \frac{2 t}{t-u} \partial_t
+\frac{u'+2}{2 (t-u)}-\frac{{u'}^2}{8 u}  
\right.}\\ \displaystyle{\left.\phantom{1}
 -\frac{1}{4}\left(
2 \alpha u +\frac{\beta}{u}
- \gamma \frac{u^3 + 4 x u^2 + 4 x^2 u}{2} 
- \delta (2 u^2 + 4 x u) 
\right)
\right\rbrack \psi_{\rm h}=0,
}\\ \displaystyle{
}
\end{array}
\right.
\label{eqLax-PIV-second-half}
\end{eqnarray}

% ============================================================ PII' Lax2h1
\begin{eqnarray}
& & {\hskip -10.0 truemm}
\PII'
\begin{array}{ll} 
\displaystyle{
%Lax2h1=
\left\lbrack
\partial_x - \frac{2}{t-u} \partial_t
+\frac{u'}{2 (t-u)}-\frac{{u'}^2}{2}  
+\alpha u +\beta\frac{u^2}{2}+\gamma (2 u^3+x u)
+ \delta \frac{u^4+x u^2}{2}
\right\rbrack \psi_{\rm h}=0.
}\\ \displaystyle{
}
\end{array}
\label{eqLax-PII-second-half}
\end{eqnarray}		
			
\vfill\eject			
% ***************************************************************** References

\vfill\eject
						
\end{document}